\documentclass[
 reprint,
 amsmath,amssymb,
 aps,
 prl,
]{revtex4-2}

\usepackage{placeins}
\usepackage{lipsum}
\usepackage{graphicx}
\usepackage{epstopdf}

\usepackage{dcolumn}
\usepackage{bm}
\usepackage[colorlinks=true, citecolor={blue!80!black}, urlcolor={blue!50!black}, linkcolor = {blue!80!black}]{hyperref}
\usepackage[colorinlistoftodos]{todonotes}
\usepackage[capitalize]{cleveref}
\usepackage[english]{babel}
\usepackage{braket}

\usepackage{siunitx}
\DeclareSIUnit{\belmilliwatt}{Bm}
\DeclareSIUnit{\dBm}{\deci\belmilliwatt}

\begin{document}
\title{Dynamical polarization of the fermion parity in a nanowire Josephson junction}
\author{J.\,J. Wesdorp\textsuperscript{1}}
\author{L. Gr{\"u}nhaupt\textsuperscript{1}}
\author{A. Vaartjes\textsuperscript{1}}
\author{M. Pita-Vidal\textsuperscript{1}}
\author{A. Bargerbos\textsuperscript{1}}
\author{L.\,J. Splitthoff\textsuperscript{1}}
\author{P. Krogstrup\textsuperscript{3}}
\author{B. van Heck\textsuperscript{2}}
\author{G. de Lange\textsuperscript{2}}
\affiliation{\textsuperscript{1}QuTech and Kavli Institute of Nanoscience, Delft University of Technology, 2628 CJ, Delft, The Netherlands \\
\textsuperscript{2}Microsoft Quantum Lab Delft, 2628 CJ, Delft, The Netherlands \\
\textsuperscript{3} Center for Quantum Devices, Niels Bohr Institute, University of Copenhagen
and Microsoft Quantum Materials Lab Copenhagen, Denmark}

\date{\today}

\begin{abstract}
 Josephson junctions in InAs nanowires proximitized with an Al shell can host gate-tunable Andreev bound states.
 Depending on the bound state occupation, the fermion parity of the junction can be even or odd. 
Coherent control of Andreev bound states has recently been achieved within each parity sector, but it is impeded by incoherent parity switches due to excess quasiparticles in the superconducting environment.
 Here, we show that we can polarize the fermion parity dynamically using microwave pulses by embedding the junction in a superconducting LC resonator. 
We demonstrate polarization up to $94\%\pm 1\%$ ($89\%\pm 1\%$) for the even (odd) parity as verified by single shot parity-readout. Finally, we apply this scheme to probe the flux-dependent transition spectrum of the even or odd parity sector selectively, without any post-processing or heralding.
\end{abstract}

\maketitle
Josephson junctions (JJs)  
play an essential role in the field of circuit quantum electrodynamics (cQED)~\cite{blais_cavity_2004}, providing the non-linearity required for quantum-limited amplification and quantum information processing~\cite{Devoret_2013, Roy_2016, Wendin_2017, Kjaergaard_2020}. 
Microscopically, the supercurrent in JJs is carried by Andreev bound states (ABS)~\cite{kulik_macroscopic_1969, beenakker_universal_1991}. 
Recent advances in hybrid circuits, where the JJs consist of superconducting atomic break junctions~\cite{bretheau_exciting_2013, bretheau_localized_2013, janvier_coherent_2015} or  superconductor-semiconductor-superconductor weak links~\cite{de_lange_realization_2015, larsen_semiconductor-nanowire-based_2015, casparis_superconducting_2018, pita-vidal_gate-tunable_2020}, have opened up exciting new avenues of research due to the few, transparent, and tunable ABS that govern the supercurrent. 

ABS are fermionic states that occur in Kramers degenerate doublets~\cite{beenakker_universal_1991}.
Their energy depends on the phase difference across the JJ, and the degeneracy can be lifted in the presence of spin-orbit coupling~\cite{chtchelkatchev_andreev_2003} or a magnetic field. 
Each doublet can be occupied by zero or two, or one quasiparticle (QP), giving rise to even and odd parity sectors.
Theoretical proposals have investigated both sectors as qubit degrees of freedom~\cite{zazunov_andreev_2003, desposito_controlled_2001, chtchelkatchev_andreev_2003, padurariu_theoretical_2010}, relying on conservation of parity.
Although fermion parity is conserved in a closed system, superconducting circuits are known to contain a large non-equilibrium population of QPs ~\cite{glazman_bogoliubov_2021,  aumentado_nonequilibrium_2004, lenander_measurement_2011, sun_measurements_2012,riste_millisecond_2013,wenner_excitation_2013,pop_coherent_2014,vool_non-poissonian_2014,wang_measurement_2014,riwar_normal-metal_2016,serniak_hot_2018,uilhoorn_quasiparticle_2021}. 
These QPs can enter the junction and ``poison'' the ABS on timescales of $\approx \SI{100}{\micro\second}$~\cite{zgirski_evidence_2011, hays_direct_2018, janvier_coherent_2015}.
Recent experiments exploring ABS dynamics in cQED architectures have shown remarkable control over the ABS using microwave drives.
Refs.~\cite{janvier_coherent_2015, hays_direct_2018} were able to demonstrate coherent manipulation in the even parity manifold, while Refs.~\cite{tosi_spin-orbit_2019, hays_continuous_2020, hays_coherent_2021} focused on the odd manifold and could coherently control a trapped QP and its spin.
Both schemes have to monitor random poisoning events in order to operate in the intended parity sector. 

So far, the route to controlling the ABS parity has been by engineering the free energy landscape via electrostatic~\cite{ van_dam_supercurrent_2006, de_franceschi_hybrid_2010} or flux~\cite{zgirski_evidence_2011} tuning such that the equilibrium rates of QP trapping and de-trapping become strongly unbalanced.
Practical applications, like Andreev qubits~\cite{zazunov_andreev_2003, desposito_controlled_2001, chtchelkatchev_andreev_2003, padurariu_theoretical_2010, park_andreev_2017, janvier_coherent_2015, hays_direct_2018, hays_coherent_2021} or Majorana detection~\cite{prada_andreev_2020} for topological qubits~\cite{karzig_scalable_2017}, require to dynamically set the parity without changing gate or flux settings - e.g. using a microwave drive. 
In a closed system, microwave photons are only allowed to drive transition that preserve parity.
Nevertheless, a microwave drive should be able to polarize the fermion parity locally, at the junction, by driving transitions that end up exciting one QP into the continuum of states above the superconducting gap in the leads~\cite{chtchelkatchev_andreev_2003, riwar_shooting_2014,  klees_nonequilibrium_2017, olivares_dynamics_2014, riwar_control_2015}.
However, so far microwaves have only been observed to increase the rate of QP escape from the junction~\cite{levenson-falk_single-quasiparticle_2014, farmer_continuous_2021,  hays_coherent_2021} while deterministic polarization towards either parity has not yet been demonstrated.

\begin{figure}
\includegraphics{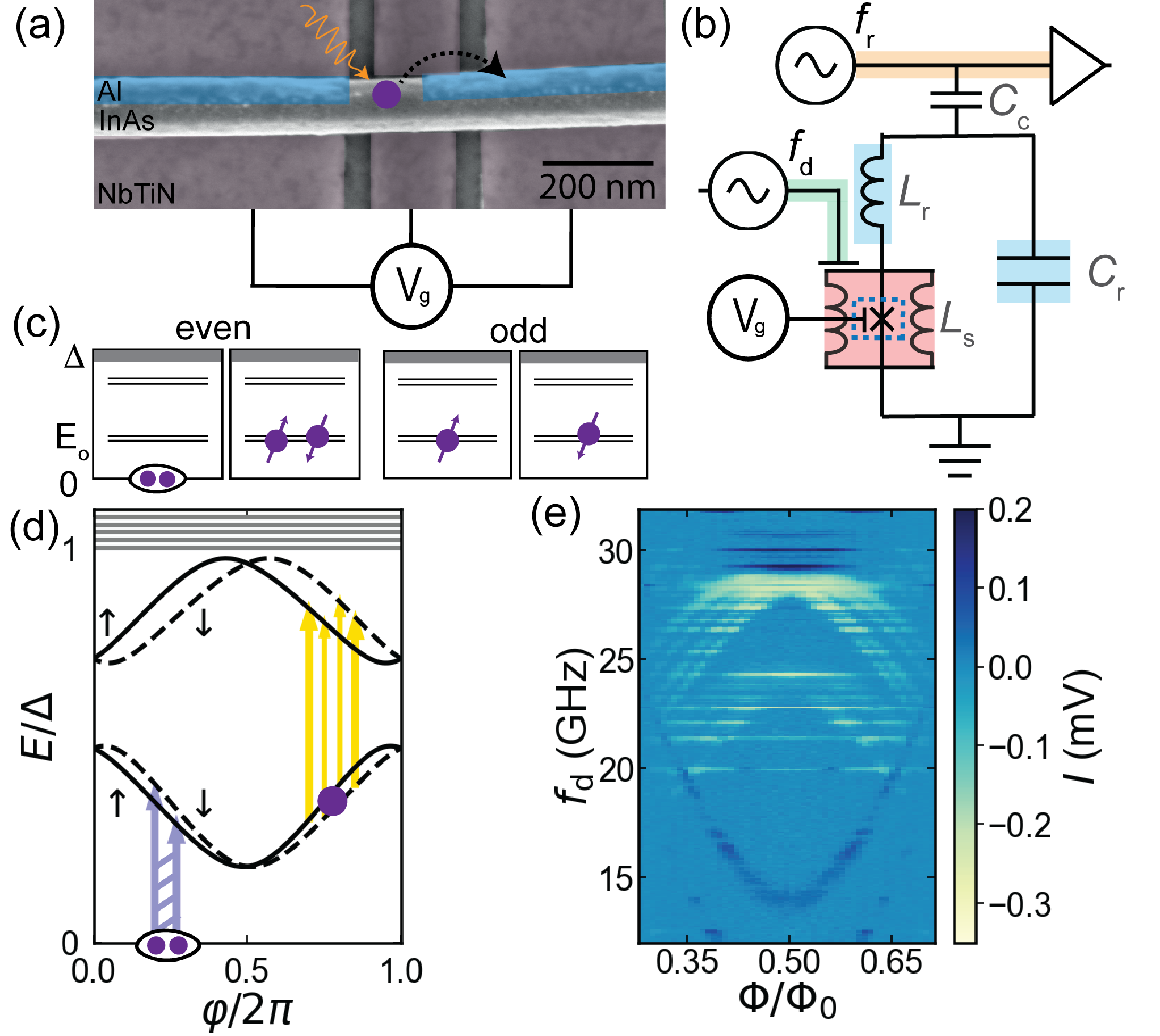}
\caption{\label{fig:f1} \textbf{(a)} False colored scanning electron micrograph\protect\footnote{Original image available in supplementary material} of the InAs/Al nanowire Josephson junction formed by etching a $\approx\SI{150}{\nano\meter}$ section of the Al shell [blue dotted box in (b)], and sketch of the detrapping of a quasiparticle (purple circle) from the junction by microwave irradiation (yellow arrow). \textbf{(b)} Setup schematic. Two parallel inductances shunt the gate-tunable nanowire Josephson junction and form a gradiometric RF SQUID (red). To allow dispersive readout of the Andreev bound state (ABS) spectrum, we integrate the SQUID into an LC resonator (blue), which is capacitively coupled to a transmission line (orange) and probed at frequency $f_\mathrm{r}$. A second transmission line (green) allows direct driving of ABS transitions via microwave tones ($f_\mathrm{d}$). \textbf{(c)} Schematic energy levels of ABS inside the superconducting gap $\Delta$ and the lowest doublet occupation configurations for even and odd junction parity. \textbf{(d)} Energy diagram of levels shown in (c) versus applied phase bias $\varphi=\Phi /\Phi_0$ tuned via an external flux $\Phi$~\cite{tosi_spin-orbit_2019}. Blue connected arrows denote the transition in the even parity sector starting from the ground state, while yellow arrows denote transitions in the odd sector starting with one of the lower levels occupied by a QP.
\textbf{(e)} Measured I-quadrature of the transmitted tone at $f_r$ versus $f_d, \Phi,$ showing the ABS spectrum containing the two types of transitions indicated in (d) based on whether the initial state had odd (yellow) or even (dark blue) parity.}
\end{figure}

In this Letter, we demonstrate dynamical polarization of the fermion parity of ABS in a nanowire Josephson junction using only microwave control.  
We first demonstrate single shot readout of the ABS parity.
We then show that we can polarize the ABS into either even or odd parity depending on the frequency and power of a second pumping tone.
Using a two-state rate model, we infer that the pumping tone can change the transition rate from even to odd parity, or vice versa, by more than an order of magnitude.
Finally we show that we can deterministically polarize the ABS parity into either even or odd over a wide range of flux by pumping at a flux-dependent frequency, and confirm this with parity selective spectroscopy without any post-selection or heralding.

We focus on the microwave transition spectrum of ABS confined to an InAs nanowire Josephson junction embedded in a radio-frequency superconducting quantum interference device (RF SQUID)~[\cref{fig:f1}(a)]~\cite{clarke_squid_2004} acting as a variable series inductance in an LC resonator tank circuit~[\cref{fig:f1}(b)]~\cite{wesdorp_supplementary_2021}.
For driving ABS transitions we include a separate transmission line that induces an AC voltage difference across the junction. The number of ABS levels, and therefore the inductance of the nanowire junction, is controlled through the field-effect by applying a voltage $V_\mathrm{g}$ to the bottom gates~\cite{doh_tunable_2005, van_woerkom_microwave_2017, goffman_conduction_2017}. 
In order to have a consistent dataset, the gate value is kept fixed throughout this Letter at $V_\mathrm{g}=0.6248$ V~\cite{wesdorp_supplementary_2021}.

At this particular $V_\mathrm{g}$, ABS transitions are visible using standard two-tone spectroscopy~[\cref{fig:f1}(e)] in the flux range between $0.3 \Phi_0$ and 0.7$\Phi_0$. Where $\Phi_0=h/2e$ is the magnetic flux quantum. 
Due to a finite population of QPs in the environment~\cite{glazman_bogoliubov_2021}, in the absence of any drive, the parity of the junction fluctuates during the measurement~\cite{hays_direct_2018, janvier_coherent_2015, zgirski_evidence_2011}.
As a consequence, the measured transition spectrum~[\cref{fig:f1}(e)] is the sum of two sets of transitions with an initial state of either even or odd parity with opposite response in the I-Q plane~\cite{metzger_circuit-qed_2021,park_adiabatic_2020}.
In~\cref{fig:f1}(c) we depict a schematic~\cite{wesdorp_supplementary_2021} of the relevant ABS levels for this particular $V_\mathrm{g}$. The lowest doublet consists of two spin-dependent fermionic levels (energies $E_o^\uparrow, E_o^\downarrow)$ that can either be occupied by a QP or not ~\cite{van_heck_zeeman_2017}. 
The presence of odd-parity transitions [yellow lines in~\cref{fig:f1} (e)] requires that another doublet is present at higher energies.
These can generally be present in finite length weak links or in the presence of multiple transport channels.
The ABS levels are spin-split at zero field and finite phase drop $\varphi$ over the junction, because spin-orbit coupling induces a spin- and momentum-dependent phase shift gained while traversing the finite length weak link~\cite{governale_spin_2002, park_andreev_2017, hays_continuous_2020, tosi_spin-orbit_2019}.
This is depicted in the phase dependence of the ABS energies in~\cref{fig:f1} (d)~\cite{tosi_spin-orbit_2019, wesdorp_supplementary_2021}.
The two sets of transitions given either an odd or even parity initial state of in~\cref{fig:f1}(e) are indicated in~\cref{fig:f1}(d) by yellow or blue arrows respectively.

\begin{figure}
\includegraphics{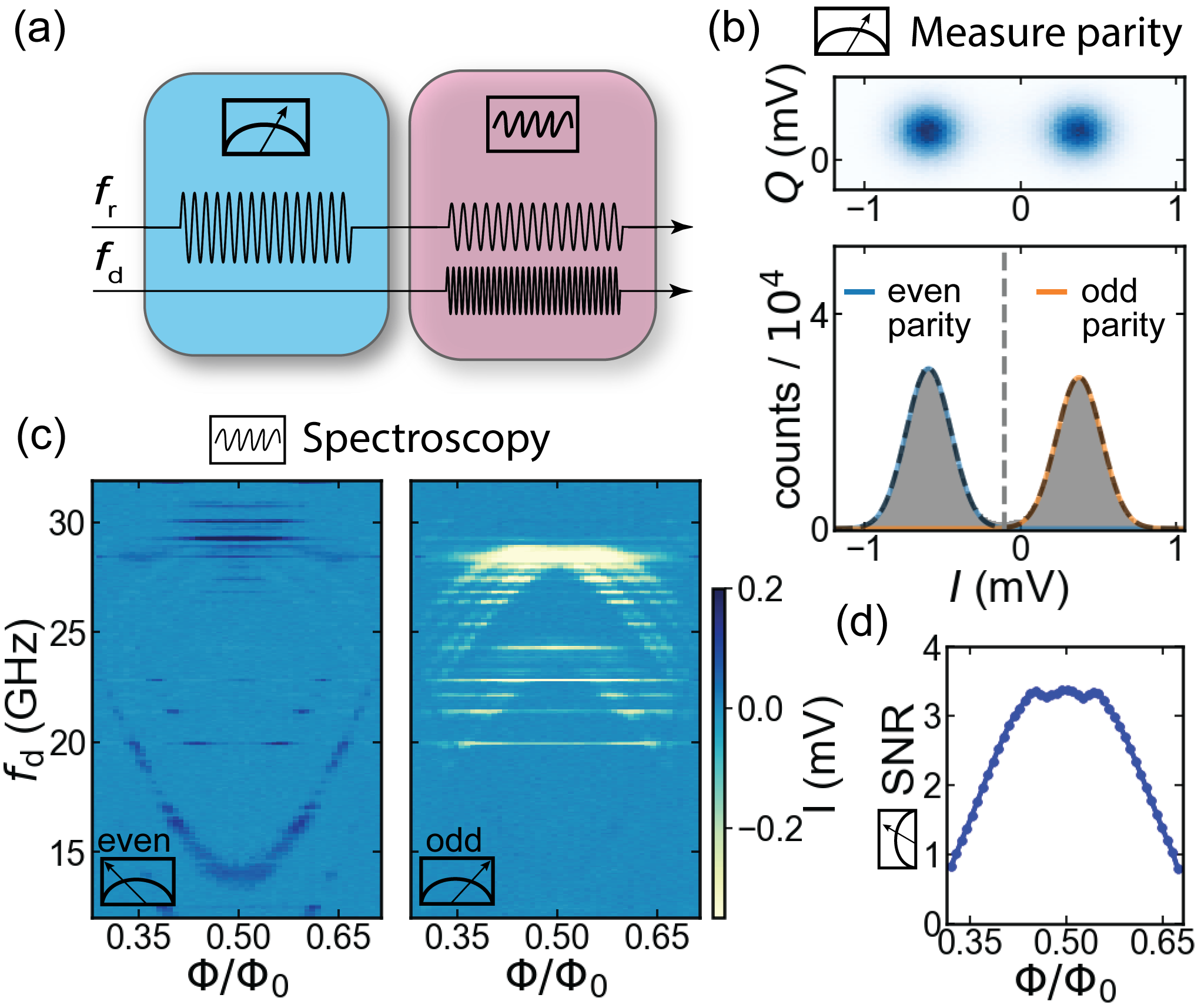}
\caption{\label{fig:fig2} Spectroscopy conditioned on the result of an initial single shot parity readout. \textbf{(a)} Pulse sequence. For each drive frequency $f_\mathrm{d}$, we first measure the initial junction parity with a strong \SI{20}{\micro\second} readout pulse at frequency $f_\mathrm{r}$ and subsequently perform ABS spectroscopy using a weaker \SI{20}{\micro\second} pulse at $f_\mathrm{r}$ together with a pulse at $f_\mathrm{d}$ on the drive line. \textbf{(b)} Top - 2D Histogram of rotated parity measurement outcomes at $\Phi=0.44\Phi_0$ in the I-Q plane. Bottom - histogram of the projection to the I-axis (grey bars) fitted to a double Gaussian distribution (dashed black line). Blue (orange) lines show single Gaussians using the previously fitted parameters indicating even (odd) initial parity. Dashed grey line indicates the threshold used for parity selection. \textbf{(c)} Post-processed spectroscopy results of the second pulse conditioned on the initial parity, i.e the first measurement being left or right from the threshold indicated in (b). This allows to separate the transition spectrum by initial state parity~[cf.~\cref{fig:f1}(e)]. \textbf{(d)} Signal to noise ratio (SNR) of the parity measurement.}
\end{figure}

In order to identify the parity of the initial state,
we perform spectroscopy conditioned on a single shot measurement of the ABS parity~[\cref{fig:fig2}(a)].
The measurement outcomes of the first pulse are distributed as two Gaussian sets in the I-Q plane corresponding to the two parities~[\cref{fig:fig2}(b)]. 
We fit a double Gaussian distribution to the projection towards the  $I$-axis (black line)~\cite{wesdorp_supplementary_2021}.
We then extract the population $p_\mathrm{e}$ ($p_\mathrm{o}$) of the lowest energy ABS with even (odd) parity via the normalized amplitudes of the fitted Gaussians depicted with a blue line (orange line)~\cite{wesdorp_supplementary_2021}.
For each $\Phi$ we determined a selection threshold $I_T$ to distinguish which parity was measured with the first pulse~\cite{wesdorp_supplementary_2021}.
We then post-select the second pulse data conditioned on having $I < I_T$ ($I > I_T)$ in the first pulse.
This allows us to verify that the two outcomes belong to the even (odd) parity branches by comparing the resulting two-tone spectra~[\cref{fig:fig2}(c)] to~\cref{fig:f1}(d).
Finally, we quantify how well we can select on parity by investigating the signal to noise ratio (SNR) of the parity measurement~[\cref{fig:fig2}(d)]~\cite{wesdorp_supplementary_2021}.
This SNR changes as a function of $\Phi$, reflecting the strong flux dependence of the dispersive shifts of the resonator corresponding to different transitions~\cite{janvier_coherent_2016, metzger_circuit-qed_2021}.

\begin{figure*}
\includegraphics{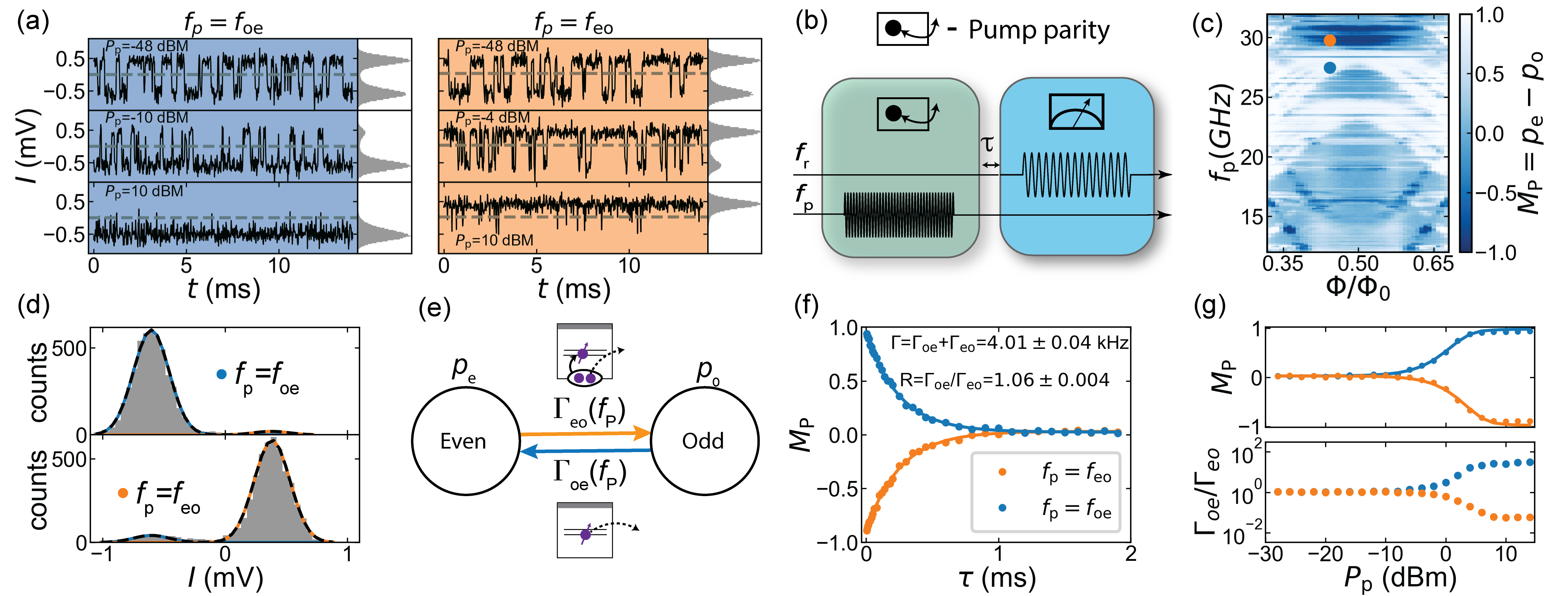}
\caption{\label{fig:fig3} Dynamic polarization of the junction parity via microwave pumping. 
\textbf{(a)} \SI{15}{\milli\second} trace showing continuous monitoring of the parity (\SI{20}{\micro\second} integration time) while applying a second tone resonant with one of the odd ($f_\mathrm{oe}=\SI{27.48}{\giga\hertz}$) or even ($f_\mathrm{eo}=\SI{29.72}{\giga\hertz}$) parity transitions at low, medium and strong drive power. Grey histograms show all measured points in the \SI{2}{\second} trace. \textbf{(b)} Pulse scheme used to verify the effect of pumping for panels (c-g). A \SI{50}{\micro\second} pumping pulse at frequency $f_\mathrm{p}$ is followed after a delay $\tau$ by the parity measurement described in~\cref{fig:fig2}. Note that in (c) a low power tone at $f_\mathrm{r}$ was present during the pumping pulse~\cite{wesdorp_supplementary_2021}. \textbf{(c)} Flux dependent pump map of measured parity polarization $M_\mathrm{P}$ versus $f_\mathrm{p}$ used for the first pulse, where +1 (-1) indicates complete polarization to even (odd) parity. \textbf{(d)} Histograms of $I$-values of the parity measurement after polarization ($P_\mathrm{p}$=\SI{14}{\dBm}, $\tau=\SI{4}{\micro\second}$) to even (odd) parity via pumping at $f_{oe}$ ($f_{eo}$). Flux set point and pump frequency are indicated by same colored dots in panel (c). \textbf{(e)} Phenomenological two state rate model used to describe the parity dynamics and polarization process. Dependent on $f_\mathrm{p}$, either the trapping rate $\Gamma_\mathrm{eo}$ or de-trapping rate $\Gamma_\mathrm{oe}$ increases from its equilibrium value. \textbf{(f)} Decay time experiment. First we fully polarize ($P_\mathrm{p}$=\SI{14}{\dBm}) the junction into even (blue dots) or odd (orange dots) parity and then vary $\tau$ before the parity measurement. Numbers indicate equilibrium parity switching rates $\Gamma_\mathrm{oe}$, $\Gamma_\mathrm{eo}$ extracted from an average of fits (solid lines) of the rate model for different $f_\mathrm{p}$~\cite{wesdorp_supplementary_2021}. \textbf{(g)} Pump power dependence of $M_\mathrm{P}$ used to extract the ratio of switching rates $R=\Gamma_\mathrm{oe}/\Gamma_\mathrm{eo}$ for fixed delay $\tau=\SI{4}{\micro\second}$. Error-bars in (f, g) are smaller than the used markers.}
\end{figure*}

We now investigate the effect of a strong drive on the  parity.
In the absence of drive, repeated parity measurements yield a near 50-50 split between even and odd~[\cref{fig:fig2}(b)], i.e. $
p_\mathrm{e} / p_\mathrm{o}=1.06$.
This can also be seen under continuous readout of the cavity at $f_\mathrm{r}$~[\cref{fig:fig3}(a, top)].
A second drive tone at a frequency $f_\mathrm{p}$ comparable to the ABS transition frequency changes this balance~[\cref{fig:fig3}(a, middle)].
Remarkably, for stronger powers the effect is to completely suppress one of the two measurements outcomes~[\cref{fig:fig3}(a, bottom)].
In order to rule out a direct effect on the parity readout by the strong drive, we continue with a pulsed experiment~[\cref{fig:fig3}(b)].
We send a pulse at frequency $f_\mathrm{p}$ to polarize the parity, followed by a parity measurement at $f_\mathrm{r}$ with the same settings as~\cref{fig:fig2}.
A delay $\tau=\SI{4}{\micro\second}$ is inserted between pulses to make sure the cavity is not populated via the drive.
This also allows most excited ABS population within a parity sector to decay back to their parity dependent ground state before the parity readout, since typical coherence relaxation processes are much faster than parity switching times~\cite{hays_direct_2018, hays_continuous_2020}.
In order to map out the frequency and flux dependence of the parity polarization, we perform a similar pulse sequence at high pump power versus flux $\Phi$ and pump frequency $f_\mathrm{p}$~[\cref{fig:fig3}(c)].
We quantify the polarization $M_\mathrm{P} = p_\mathrm{e}-p_\mathrm{o}$ via the parity population imbalance at the end of the sequence.
Comparing to~\cref{fig:fig2}, note that we pump from $p_\mathrm{e}$ to $p_\mathrm{o}$ if driving an even transition, and vice versa.  

For some frequencies the effect is to completely suppress one of the two measurements outcomes, indicating that at the end of the pulse, the junction is initialized in a given parity~[\cref{fig:fig3}(d)]. 
We now focus the investigation on pump frequencies that cause a strong parity imbalance at $\Phi=0.44\,\Phi_0$.
We reach $M_\mathrm{P} = 0.94\pm0.01$ for pumping on an odd parity transition ($f_\mathrm{p}=\SI{27.48}{\giga\hertz}$) and $M_\mathrm{P} = -0.89 \pm 0.01$ for pumping on an even parity transition ($f_\mathrm{p}=\SI{29.72}{\giga\hertz}$) - the resulting parity is opposite to the parity of the pumped transition.

We interpret the polarization to result from the effect of the drive on the parity transition rates.
To quantify this, we start by extracting the transition rates in absence of the drive.
We use a phenomenological model involving two rates $\Gamma_\mathrm{oe}$ (for QP de-trapping) and $\Gamma_\mathrm{eo}$ (for QP trapping) at which the junction switches between even  and odd ground states~[\cref{fig:fig3}(e)]~\cite{wesdorp_supplementary_2021}. 
We can estimate $\Gamma_\mathrm{oe}$ and $\Gamma_\mathrm{eo}$ by varying the delay $\tau$ between the drive and measurement pulse at the optimal drive frequencies that initialize the parity~[\cref{fig:fig3}(f)].
We find that the two rates are comparable in equilibrium, on average the ratio $R=\Gamma_\mathrm{oe}/\Gamma_\mathrm{eo}=1.06$ and the characteristic decay rate $\Gamma=\Gamma_\mathrm{oe}+\Gamma_\mathrm{eo}=4.01\pm 0.04$ \SI{}{\kilo\hertz}~\cite{wesdorp_supplementary_2021}.
The extracted equilibrium rates are found to be independent of the pump frequency $f_\mathrm{p}$ or pump power $P_\mathrm{p}$ used for polarization before the measurement, indicating that when the pump tone is off, the rates go back to their equilibrium value on timescales faster than the measurement time and delay used.

We then investigate the effect of the drive power on the transition rates, by performing the same pulse sequence as in~\cref{fig:fig3}(b), keeping $\tau=\SI{4}{\micro\second}$ but varying $P_\mathrm{p}$.
From the power dependence of $M_\mathrm{P}$ we extract $R$ versus power~[\cref{fig:fig3}(g)], by
assuming that we have reached a new steady state at the end of the pump tone~\cite{wesdorp_supplementary_2021}.
We see that the rates become strongly imbalanced, reaching $R=32\pm9$ $(R^{-1}=17\pm 2)$ for pumping at $f_\mathrm{oe}$ ($f_\mathrm{eo}$).

From Fermi's golden rule, a single photon process would result in a unity exponent of the power dependence of the rates.
However, a phenomenological fit [solid lines in~\cref{fig:fig3}(g)] indicates in general an exponent larger than one~\cite{wesdorp_supplementary_2021}.
We therefore suspect multi-photon processes are at play.  

\begin{figure}
\includegraphics{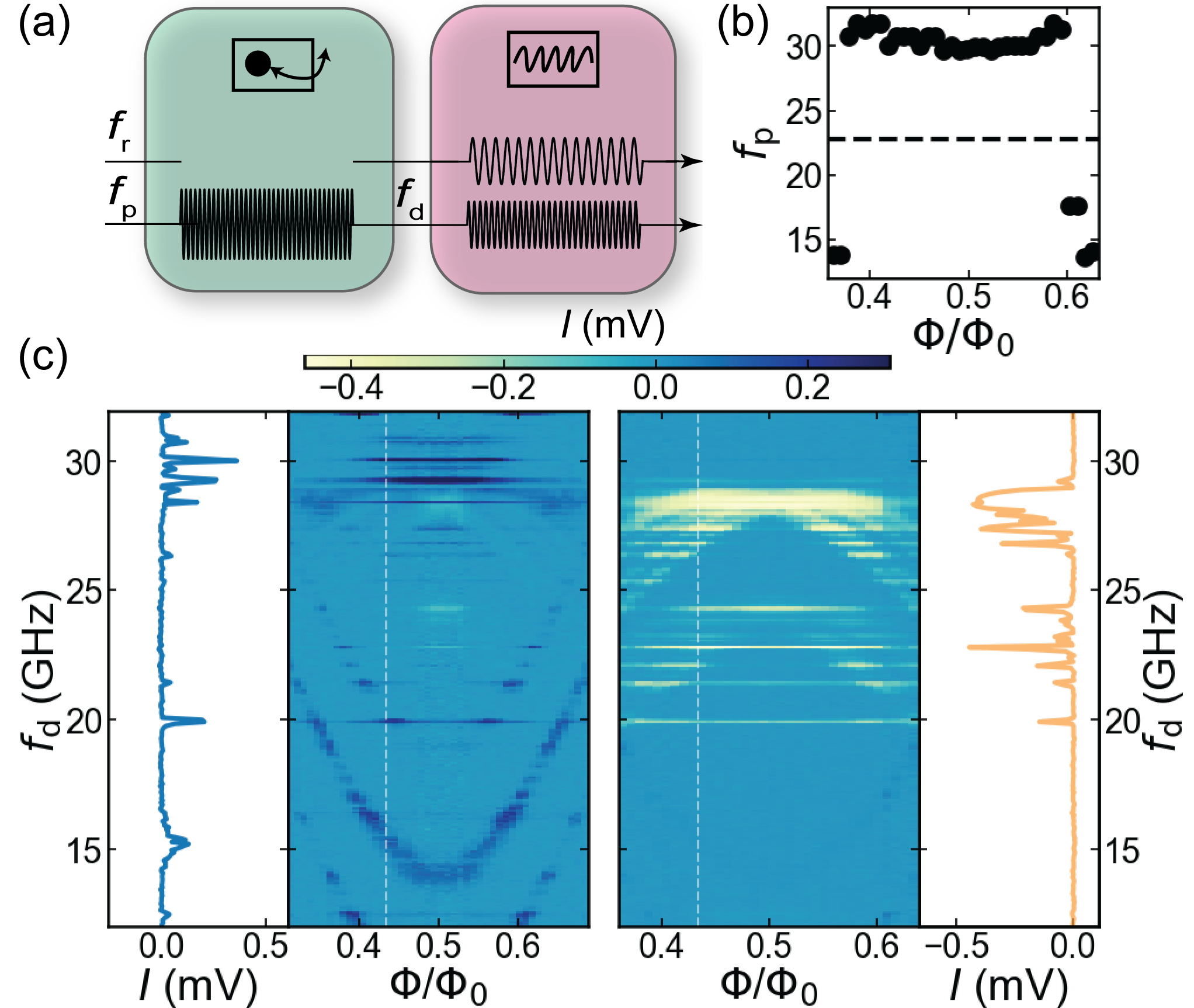}
\caption{\label{fig:fig4} Deterministic parity initialization verified by spectroscopy for a range of flux values. \textbf{(a)} Pulse sequence. We initialize the parity using a flux dependent pumping frequency for \SI{100}{\micro\second} at $f_\mathrm{p}$ together with a low power tone at $f_\mathrm{r}$. This is followed after \SI{5}{\micro\second} by a spectroscopy pulse of \SI{20}{\micro\second} similar to ~\cref{fig:f1}, but without any post-selection or heralding. \textbf{(b)} Pump frequency $f_\mathrm{p}$ used to increase $\Gamma_\mathrm{eo}$ (dots) and $\Gamma_\mathrm{oe}$ dashed line. \textbf{(c)} Result of the second spectroscopy pulse after initializing into the even parity (left panel) or odd parity (right panel). Linecuts at $\Phi=0.43\Phi_0$ demonstrate the disappearance of odd (even) transitions after initialization in even (odd) parity.}
\end{figure}
The threshold frequencies expected for the trapping and de-trapping are $\Delta$ + $\textrm{min}\{E_o^\uparrow$, $E_o^\downarrow\}$ and $\Delta - E_o^{\uparrow,\downarrow}$, respectively~\cite{riwar_shooting_2014, klees_nonequilibrium_2017,olivares_dynamics_2014,riwar_control_2015}. These thresholds correspond to the breaking of a pair into one QP in the continuum and one in the ABS, and to the excitation of a trapped QP in the continuum. However, we observe polarization at drive frequencies lower than these thresholds: $\Gamma_\mathrm{eo}$ increases already by driving at a frequency $E_o^\uparrow + E_o^\downarrow$, while $\Gamma_\mathrm{oe}$ increases when driving resonant with any odd-parity transition. We suspect this lower threshold to be due to the combination of a crowded spectrum - from the multiband-nature of our wire and other modes in the circuit~\cite{olivares_dynamics_2014} - and a strong drive. This allows ladder-like multiphoton processes, also suggested in earlier experiments~\cite{hays_coherent_2021, levenson-falk_single-quasiparticle_2014}. 
However we could not fully reconstruct the spectrum near $\Delta$, so this is only a qualitative explanation. 
Note that we also see peaks in the polarization at transitions $f_\mathrm{even,odd}\pm f_\mathrm{r}$ due to multiphoton processes involving the cavity, since a weak readout tone was on during the pumping for~\cref{fig:fig3}(c)~\cite{wesdorp_supplementary_2021}.

To demonstrate the effectiveness of the deterministic parity control, we now perform parity-selective two-tone spectroscopy without post-selection or heralding.
We deterministically initialize the parity of the junction before each spectroscopic measurement via the pumping scheme demonstrated in~\cref{fig:fig3} followed by a spectroscopy measurement~[\cref{fig:fig4}(a)].
We vary the pumping frequency $f_\mathrm{p}$ as a function of the $\Phi$ to achieve maximum polarization~[\cref{fig:fig4}(b)]. 
The optimal pumping frequency $f_\mathrm{eo}(\Phi)$ is experimentally determined from the data shown in~\cref{fig:fig3}(c). 
For $f_\mathrm{oe}(\Phi)$ we pump at a fixed frequency $f_\mathrm{oe}=22.76$ GHz because at this frequency a finite pumping rate was present for all required $\Phi$~\cite{wesdorp_supplementary_2021}.
In~\cref{fig:fig4}(c) the result is shown for even (odd) initialization on the left (right).
The similarity with the post-selected results of~\cref{fig:fig2} shows the success of the method. 

In summary, we have demonstrated deterministic polarization of the fermion parity in a nanowire Josephson junction using microwave drives.
For pumping towards even parity we can explain the maximal polarization to be limited by parity switches during the measurement pulse~\cite{wesdorp_supplementary_2021}. 
This mechanism is not sufficient to account for the higher residual infidelity when polarizing to odd parity, which we suspect is due to a finite pumping rate towards the even sector during the pump pulse. 

These results enable  fast initialization of ABS parity and thus provide a new tool for studying parity switching processes, highly relevant for Andreev ~\cite{hays_coherent_2021,janvier_coherent_2015,hays_direct_2018} and topological~\cite{karzig_scalable_2017} qubits. 

\begin{acknowledgments}
We would like to thank Ruben Grigoryan for the PCB and enclosure design and Leo Kouwenhoven for support on the project and for commenting on the manuscript. This work is part of the research project ‘Scalable circuits of Majorana qubits with topological protection’ (i39, SCMQ) with project number 14SCMQ02, which is (partly) financed by the Dutch Research Council (NWO). 
It has further been supported by the Microsoft Quantum initiative.

\textbf{Author contributions}
JJW, AV,  LJS, MPV contributed to sample fabrication and inspection. JJW, LG, AV contributed to the data acquisition and analysis with input from GdL, BvH, AB, MPV. JJW, LG, AV, wrote the manuscript with comments and input from GdL, BvH, AB, LJS, MPV. Nanowires were grown by PK. Project was supervised by GdL, BvH. 
\end{acknowledgments}
\bibliography{bibliography2}
\appendix

\end{document}


\title{Supplementary information: Dynamical polarization of the fermion parity in a nanowire Josephson junction}

\author{J.\,J. Wesdorp\textsuperscript{1}}
\author{L. Gr{\"u}nhaupt\textsuperscript{1}}
\author{A. Vaartjes\textsuperscript{1}}
\author{M. Pita-Vidal\textsuperscript{1}}
\author{A. Bargerbos\textsuperscript{1}}
\author{L.\,J. Splitthoff\textsuperscript{1}}
\author{P. Krogstrup\textsuperscript{3}}
\author{B. van Heck\textsuperscript{2}}
\author{G. de Lange\textsuperscript{2}}
\affiliation{\textsuperscript{1}QuTech and Kavli Institute of Nanoscience, Delft University of Technology, 2628 CJ, Delft, The Netherlands \\
\textsuperscript{2}Microsoft Quantum Lab Delft, 2628 CJ, Delft, The Netherlands \\
\textsuperscript{3} Center for Quantum Devices, Niels Bohr Institute, University of Copenhagen
and Microsoft Quantum Materials Lab Copenhagen, Denmark}

\maketitle
\onecolumngrid
\tableofcontents
\newpage
\newpage
\section{Methods}\label{sec:app:methods}
\subsection{Fabrication}\label{sec:app:methods:fabrication}
\begin{figure}[b]
\includegraphics{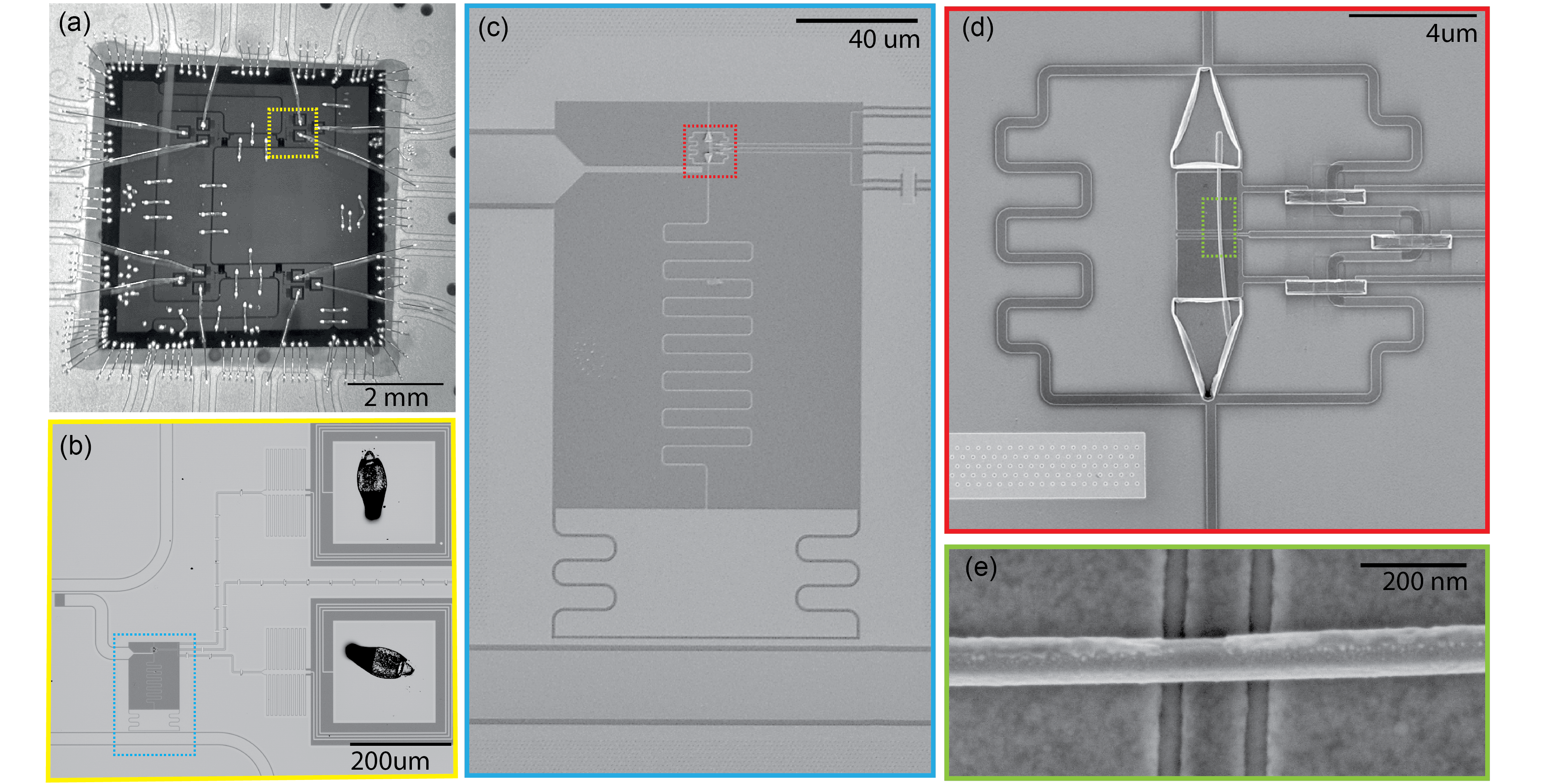}
\caption{\label{fig:sup:device-sem} Additional images of the measured device. (a, b) The chip contains four devices of which one was fully functional and studied in this work. Readout was performed for all devices via a single transmission line. A second transmission line is coupled to all devices via coupling elements acting as an effective capacitance to the device. This allows using a single line to drive multiple devices. The gate lines had on-chip LC filters to reduce high frequency noise~
\cite{mi_circuit_2017-3}. (c-e) Additional scanning electron micrographs of the device described in Fig.1 of the main text. In a $\SI{21}{\micro\meter}$ radious around the resonator, as well as in the capacitor plate, transmission-lines and drive-line, \SI{80}{\nano\meter} diameter round vortex pinning sites were patterned to reduce flux jumps and vortex induced losses when applying magnetic fields~\cite{kroll_magnetic_2019-1}. Furthermore the ground plane was patterned with $\SI{500}{\nano\meter}$ square holes to trap residual flux.}
\end{figure}

The whole circuit~[\cref{fig:sup:device-sem}] is patterned in a sputtered \SI{22}{\nano\meter} thick NbTiN film  with a kinetic inductance of around \SI{11}{\pico\henry}/square using SF6/O2 reactive ion etching. Subsequently, \SI{28}{\nano\meter} of $\mathrm{Si}_3\mathrm{N}_4$ is deposited using plasma-enhanced chemical vapor deposition (PECVD) at 300 $^{\circ}$C and patterned using a 3 minute 20:1 BOE (HF) dip with surfactant, serving both as a bottom gate-dielectric and as isolation for \SI{75}{\nano\meter} sputtered NbTiN bridges connecting the separated ground plane around the gate lines. The hexagonal nanowire has a diameter of $\approx$ \SI{80}{\nano\meter} and is epitaxially covered~\cite{chang_hard_2015} on 2 facets by a \SI{6}{\nano\meter} Al shell~[Fig. 1(a)]. It is transfered using a nanomanipulator on top of a NbTiN gate structure separated by  $\mathrm{Si}_3\mathrm{N}_4$  dielectric. The $\approx\SI{150}{\nano\meter}$ junction is etched with a 55 second MF321 (alkaline) etching step. Finally the nanowires are contacted by \SI{150}{\nano\meter} of sputtered NbTiN after 3 min of in-situ AR-milling at 50W. 

\subsection{Circuit design}\label{sec:app:methods:circuit-details}
\begin{figure}[ht]
\includegraphics{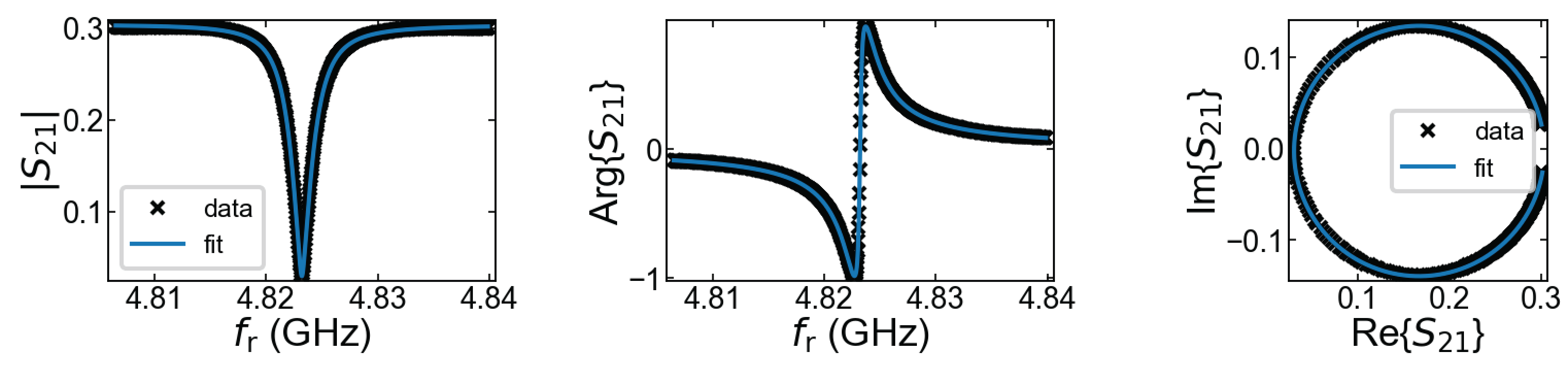}
\caption{\label{fig:sup:resonator-fit} Measurement and fit of resonance of the readout resonator shown in~\cref{fig:sup:device-sem}.  }
\end{figure}

The circuit shown in Fig.1(b) consists of a lumped element readout resonator with a resonance frequency  $f_\mathrm{c}=4.823$GHz ($L_r\approx \SI{21}{\nano\henry},L_s\approx \SI{0.7}{\nano\henry}$, $C_r\approx \SI{47}{\femto\farad}$), which is overcoupled to a \SI{50}{\ohm} transmission line.  A chosen $C_c\approx \SI{4}{\femto\farad}$ results in a coupling quality factor $Q_c= 1.7 \cdot 10^3$.  The coupling and internal quality factor $Q_i= 15  \cdot 10^3$ are extracted using the model in~\cite{khalil_analysis_2012} for a fit at average intra-cavity photon number $\langle n_\mathrm{ph}\rangle \approx 1800$ 
~\cite{bruno_reducing_2015} as shown in~\cref{fig:sup:resonator-fit} (a fit at $\langle n_{ph}\rangle \approx 17$ gave similar results).  
Typical coupling to the ABS was designed to be $g/h=I_s\frac{L_s}{L_s+L_r}\sqrt{\frac{\hbar Z_{Lc}}{2}}\approx \SI{250}{\mega\hertz}$ at $\varphi=\pi$ using a single channel ABS model~\cite{zazunov_andreev_2003} with $I_s\approx\SI{10}{nA}$. Note that the actual coupling strength depends on flux and $I_s$, which also depends on $V_g$.
We set $\Phi$ using a magnetic field with a vector magnet applied perpendicular to the nanowire but in plane with the NbTiN film, to reduce flux jumps. The effective loop area then consists of twice the area $A$ under the nanowire between the contacts due to the gradiometric design~\cite{wesdorp_preparation_2021}. The field corresponding to one flux period is $\SI{3.65}{\milli\tesla}$ ($A=\SI{0.28}{\micro\meter^2}$). By choosing $L_s \ll L_j$, we ensure that the phase drop $\varphi$ over the junction is proportional to the external flux threading the loop $\varphi=2\pi\Phi/\Phi_0$~\cite{clarke_squid_2004}.

\begin{figure*}
\includegraphics[width=0.8\textwidth]{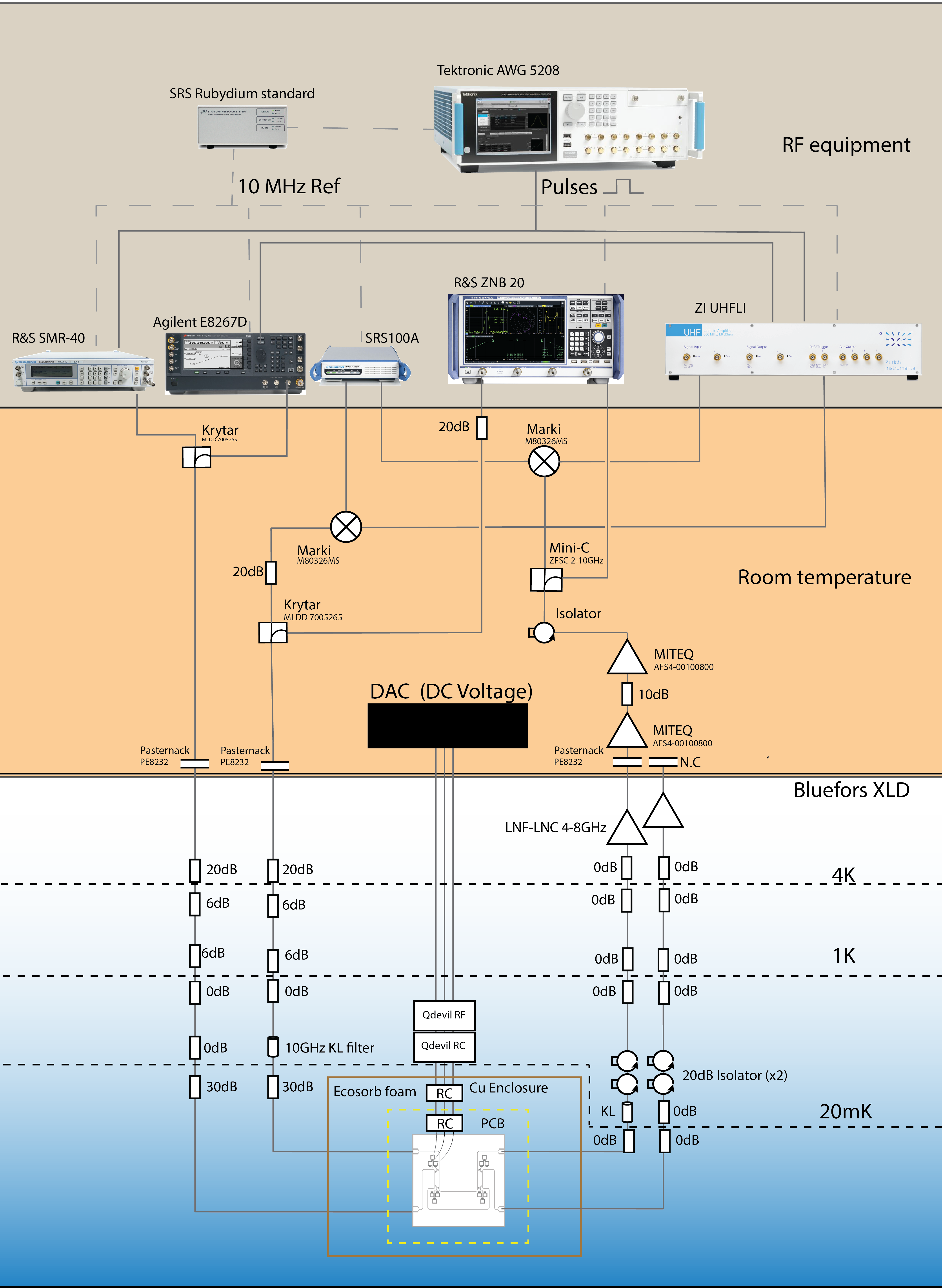}
\caption{\label{fig:sup:wiring-diagram} Full wiring diagram of the experiment. }
\end{figure*}

\subsection{Wiring diagram}\label{sec:app:methods:wiring-diagram}
A wiring diagram is shown in~\cref{fig:sup:wiring-diagram}. A R\&S ZNB 20 VNA was combined with a standard homodyne detection circuit using a splitter. 
A Zurich instrument high frequency lockin amplifier (UHFLI) both generated and demodulated a microwave tone between 500 MHz and 600 MHZ using the same internal oscillator.
This signal was upconverted by mixing it with the RF output of a R\&S SRS 100A microwave source set to a fixed frequency of 4237.11 MHz resonant with the frequency of another resonator on the chip to minimize LO leakage. 
After traveling through the fridge, the signal was amplified at 4K using a LNF 4-8 GHz HEMT amplifier as well as by two amplifiers at room temperature. This was then downconverted by mixing with the LO output of the SRS100A microwave source and demodulated in the UHFLI to obtain the I,Q values shown in the main text. All RF instruments were synced using a 10MHz rubydium reference. 

Pulse sequences were generated on both a Tektronic AWG 5208 and the internal AWG of the UHFLI and were both set to a clock frequency of 1.8 GHz. The UHFLI AWG was set to a sampling frequency of 225MHz. 
The Tektronic AWG send pulses (square) on 2 channels:
\begin{enumerate}
    \item A first long pulse gated UHFLI data streaming -allowing for a low duty-cycle measurement to circumvent ethernet bandwidth problems when streaming at a UHFLI sampling rate of $\approx$1 MHZ. The same pulse triggered the UHFLI internal AWG to start.
    \item A second pulse controlled the R\&S SMR drive pulse-modulation used to pump parity.  
\end{enumerate}
The UHFLI internal AWG also send two pulse sequences:
\begin{enumerate}
    \item The first sequence amplitude modulated the internal oscillator output of the UHFLI 
    \item the second sequence was send to the pulse-modulation input of the Agilent E8267D microwave source used for the spectroscopy drive. 
\end{enumerate}
The roughly \SI{150}{\nano\second} delay due to activation of pulse modulation and fridge traveling time were calibrated out using the internal scope function of the UHFLI. 
Readout amplitudes $A$ quoted in this work correspond to $A=V_\mathrm{pp}/1.5 \mathrm{V}$ of the carrier sine wave at $f_\mathrm{r}$ used for readout pulses.

\subsection{Measurement methods}\label{sec:sup:measurement_methods}
For all two tone spectroscopy data in the paper, we determined the optimal readout point by fitting a simple Lorentzian to $|S_{21}|$~[\cref{fig:sup:resonator-fit} (a)] by taking a frequency dependence for each $\Phi$ value. 
We then took the readout point to be at the minimum of this fitted Lorentzian. We found fitting a single Lorentzian worked even in the case of a split resonator from the dispersive shift of the even state close to $\Phi=0.5\Phi_0$. This resulted in measuring in the middle between the even/odd shifted resonator, allowing for both even and odd parity readout.

\subsection{Gate operation point}\label{sec:sup:gate_operation}
\begin{figure}
\includegraphics{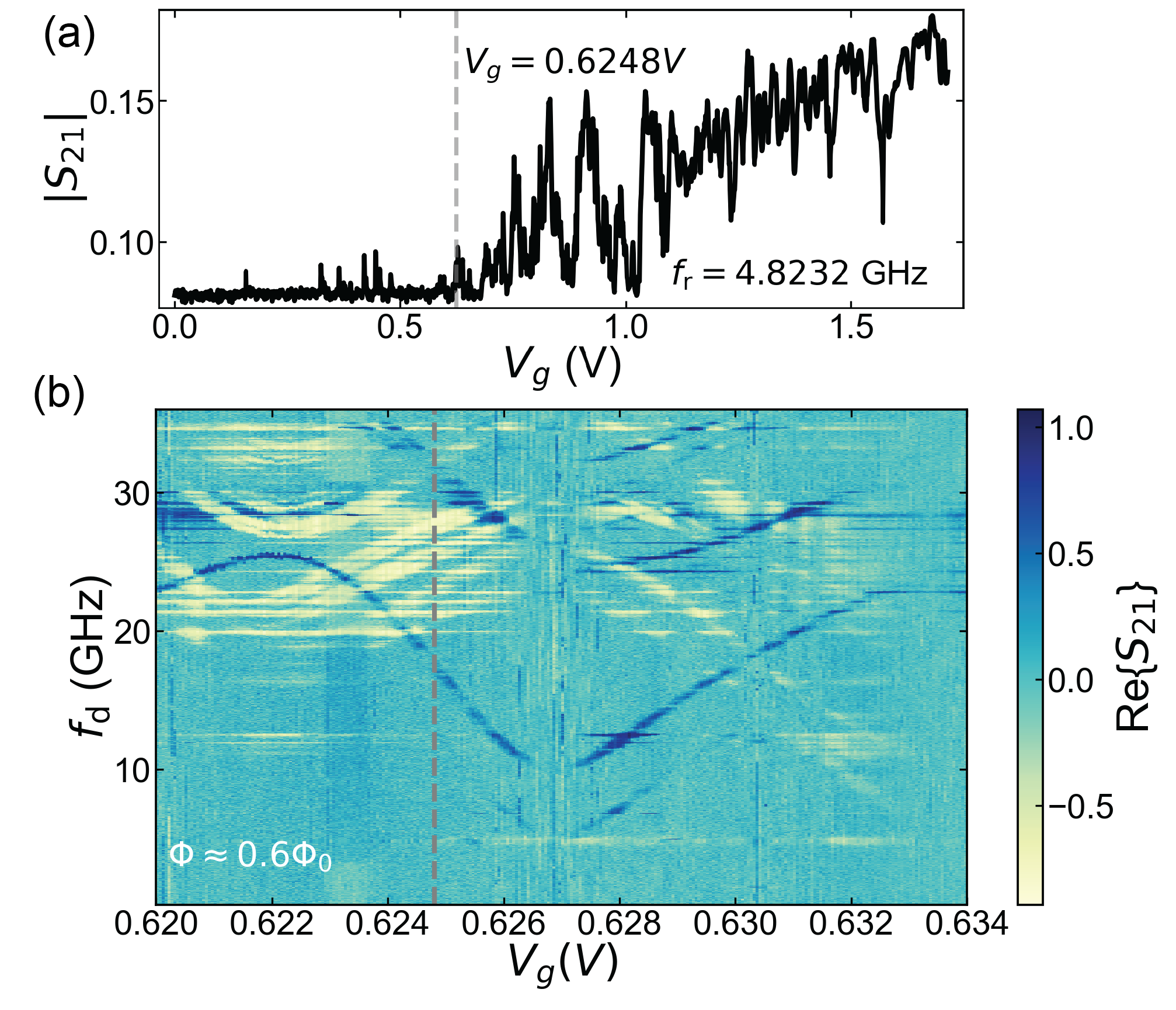}
\caption{\label{fig:sup:gate-dependence} (a) Pinch-off trace, monitoring the magnitude $|S_{21}|$ of the transmitted signal at fixed frequency indicated versus applied gate voltage. Also indicated is the operating point $V_g=0.6248$ for all other data taken.
(b) Two tone trace taken at $\Phi\approx0.6\Phi_0$ showing the ABS dispersion versus a small gate range. Grey dashed line indicate again the operating point. }
\end{figure}

In~\cref{fig:sup:resonator-fit}(a) we show an RF version of a typical pinch-off trace of the supercurrent. Here we monitor the magnitude of the transmitted signal, which is a proxy for a change in resonance frequency $f_c$ of the resonator.  At $0.5\Phi_0$ as shown here, $f_c$ goes up when the magnitude of the supercurrent increases - or similarly the inductance decreases. The trace shows mesoscopic oscillations as often seen in these systems~\cite{doh_tunable_2005}, but nevertheless has an increasing trend with $V_\mathrm{g}$. We stay close to pinch-off such that we stay in the few-mode regime as shown in a two tone trace at $\Phi=0.6\Phi_0$~[\cref{fig:sup:resonator-fit}(c)].
The gate dependence of odd and even states show a clear opposite trend~\cite{tosi_spin-orbit_2019}, since when the transparency of ABS decreases, the interband odd transitions go down in frequency while the even transitions go up.
As described in the main text, for all data taken in the main text figures we kept the gate voltage fixed at $V_g=0.6248$ V during the three weeks of data taking for this experiment in order to have a consistent dataset. $V_g$ was chosen to minimize overlap between even and odd parity transitions in the spectrum. The lowest available transition was taken to be far away from the resonator to prevent non-linearities in the cavity at high $n_\mathrm{ph}$ and facilitate parity readout. 
We expect the polarization to be possible also at other gate voltages where ABS transitions are available. However, since we suspect the polarization is caused by ladder-like processes, it might be that the polarization becomes harder (easier) when the spectrum is less (more) crowded.

\section{Data analysis and additional information}
In general, for all figures, the measured I-Q values during a parity or spectroscopy pulse pulse give two Gaussian distributed sets of outcomes in the I-Q plane~[see e.g. Fig. 2(b)]. These are rotated to maximize the variance in the I-quadrature.  Subsequently they are projected towards $I$ for each $\Phi$ separately, since the readout frequency is $\Phi$ dependent. For 2D measured spectra(c.f Fig 1., Fig 2, Fig 4.) we also subtracted a flux-dependent background - the median of $I$ of all $f_d$ for each $\Phi$ - to compensate for the change in $f_r(\Phi)$.
 
\subsection{Parity selective spectroscopy - Fig.1, Fig.2}
We now describe the analysis steps used to create the results of~Fig. 1.(e) and Fig. 2.
For the spectroscopy data of Fig. 1 in the main text, we used the average of all shots of the measured data in the spectroscopy pulse for Fig. 2 - e.g without any post-selection.
For the pulse sequence used in Fig.2 (a), we first sent a \SI{20}{\micro\second} readout pulse at frequency $f_\mathrm{r}$  ($A=0.05$), followed by a \SI{20}{\micro\second} two-tone spectroscopy sequence, i.e reading out at $f_\mathrm{r}$ ($A=0.025$) while driving at $f_\mathrm{d}$ ($P_\mathrm{d}=\SI{10}{dBm})$. 
To empty the cavity between parity measurement and spectroscopy, we inserted a \SI{5}{\micro\second} wait time.
Note that the drive power is sufficient to also induce parity pumping (see ~\cref{fig:sup:pump_length}) in addition to exciting the transition directly. We found this to increase the contrast in the spectrum. 
The sequence was repeated every \SI{1.2}{\milli\second} for each shot, in order to make sure the junction returned to its equilibrium state. This also made sure any parity pumping in the spectroscopy did not affect the subsequent shot. 
From the total line attenuation and adding $\approx\SI{6}{\deci\bel}$ loss due to the skin-effect and insertion losses, we estimate average photon number in the resonator during parity readout to be $\langle n_\mathrm{ph}\rangle\approx 44$ ($P_\mathrm{in}=-118$ dBm) and $\langle n_\mathrm{ph}\rangle\approx 11$  ($P_\mathrm{in}=-124$ dBm) during the spectroscopy pulse~\cite{bruno_reducing_2015}.

For the parity readout pulse, the rotated 1D histograms of $I$ are fitted to a double Gaussian distribution (black line in~Fig. 2) of the form $c(x) = a_1 /\sqrt{2\pi\sigma_1^2} \exp(-(x - \bar{x_1})^2/2\sigma_1^2) +  a_2 /\sqrt{2\pi\sigma_2^2} \exp(-(x - \bar{x_2})^2/2\sigma_2^2)$. 
For each $\Phi$ we determined a selection threshold $I_T(\Phi) = \left(F^{-1}(0.4)+F^{-1}(0.6)\right)/2$ where $F^{-1}$ denotes the inverse function of the cumulative normalized histogram of measured $I$ values.
The threshold for $\Phi=0.44\Phi_0$ is indicated in~Fig. 2(b).
We then post-select the data of the second pulse conditioned on having $I < I_T$ ($I > I_T)$ in the first pulse, keeping all data.
Note that it is possible to improve the accuracy of the selection if we selected further away from the threshold, keeping less data.
We define the signal-to-noise ratio as $\mathrm{SNR}=|\bar{x}_e-\bar{x}_o|/2\sigma$~\cite{de_jong_rapid_2019}. 
Here, $\bar x_e$($\bar x_o)$ is the mean of the fitted Gaussian belonging to the even (odd) parity and $\sigma$ the standard deviation, which is kept fixed to the values found in~[Fig. 2(b)] and kept the same for both Gaussians. Note that for $0.46\Phi_0<\Phi<0.54\Phi_0$ we see a slight deviation from the fit, reducing the validity of the SNR estimate, possibly due to a small readout-induced excited population. Letting $\sigma$ free as a fit parameter then results in a maximally 8\% reduction in the extracted SNR.
Extracting $R$ versus $\Phi$ from the fitted amplitudes resulted in a $approx$0.1 variation in $R$ over the flux range. 

\subsection{Pulsed polarization measurements - Fig.3}

We now describe the procedure used to obtain the data in~Fig. 3(d-g) using the pulse sequence of Fig. 3(b). For this data, we varied the pump power $P_\mathrm{p}$ for four pump frequencies $f_\mathrm{p}$. This was repeated for each delay time $\tau$. An additional \SI{1}{\milli\second} of waiting time was introduced after the two-pulse sequence to get back to equilibrium before the next sequence. 
For each $f_\mathrm{p}$, we do a double Gaussian fit to the rotated $I$ histograms of the 2nd pulse measurement shots for all $P_\mathrm{p}$ together to obtain a single $\sigma$ and two means $x_1$, $x_2$.
We then keep the means and the single $\sigma$ fixed for each $\tau$. Measuring each $\tau$ versus power took about 30 minutes so we allowed for a small variation in the mean of the Gaussians due to slow drift in the setup. We fit $a_1$ and $a_2$ for each $P_\mathrm{p}$. We then obtain the populations by normalizing $p_\mathrm{o} = \frac{a_1}{a_1+a_2}, p_\mathrm{e}=\frac{a_2}{a_1+a_2}$.
Uncertainties in $M_\mathrm{P}$ and $R$ follow from propagating the error in the fit uncertainties of $a_1, a_2$.

\begin{figure}[h]
    \centering
    \includegraphics{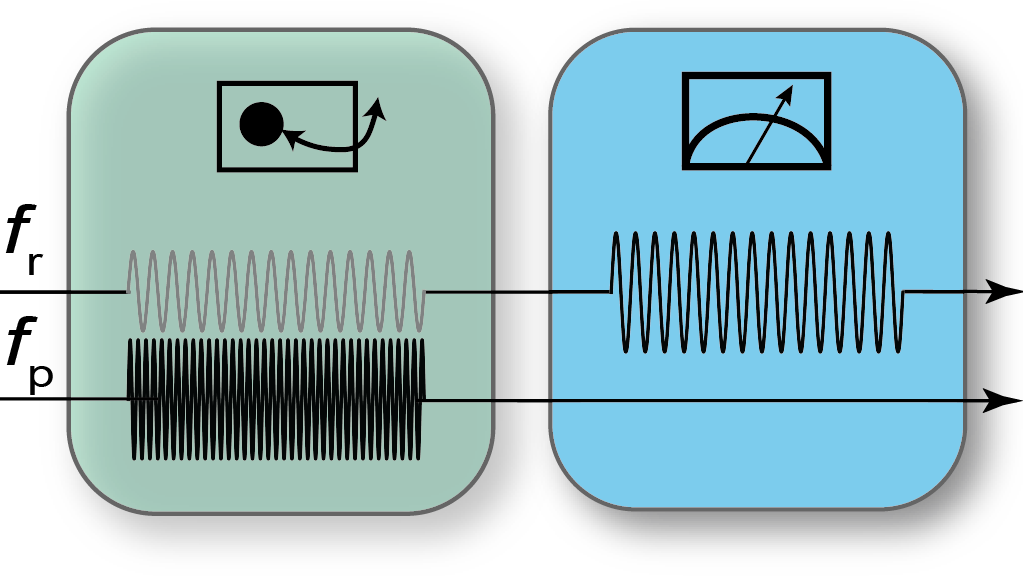}
    \caption{Pulse sequence used in Fig. 3(c)}
    \label{fig:sup:pulse_sequence_3c}
\end{figure}

The pump-frequency map of Fig.3 (c) was analysed similarly as described above, but keeping $\sigma$ fixed at all fluxes. By inspection of the fits and the residuals $\chi^2$, for some drive frequencies the fit residuals were very large (e.g. the data did no longer match a double Gaussian), resulting in horizontal lines in the plot. These we attribute to circuit resonances affecting the readout when excited with the drive tone. 
As stated in the main text, the second pulse (parity readout) was the same used for Fig. 2, Fig. 3(d-g). However, for the  pumping pulse, next to a tone at $f_\mathrm{p}$, a second weak tone at $f_r(A=0.02)$ was present (see  in~\cref{fig:sup:pulse_sequence_3c}), which we expected to help the pumping (see discussion in~\cref{sec:sup:discussion-power-dependence})). 
The wait time after each shot was reduced to \SI{200}{\micro\second} in order to save measurement time for the large 2D map, which is also the case for Fig. 4.   

\subsection{Deterministic parity initialization spectroscopy - Fig. 4}
For this dataset we used~Fig. 3 to estimate the best pumping frequency for the polarization emperically for $f_\mathrm{eo}(\Phi)$, by looking where $M_\mathrm{P} (\Phi, f_\mathrm{p})$ was maximal. We then applied the pulse sequence described in Fig. 4, for the even and odd initialization separately. Note that for $f_\mathrm{oe}(\Phi)$ we pumped at a fixed frequency $f_\mathrm{oe}=22.76$ GHz, because the crowded spectrum of odd transitions there gave a finite pumping rate over the whole required flux range.  
\newpage
\newpage
\section{Comparison of measured spectrum to theory}
\begin{figure}[ht]
\includegraphics{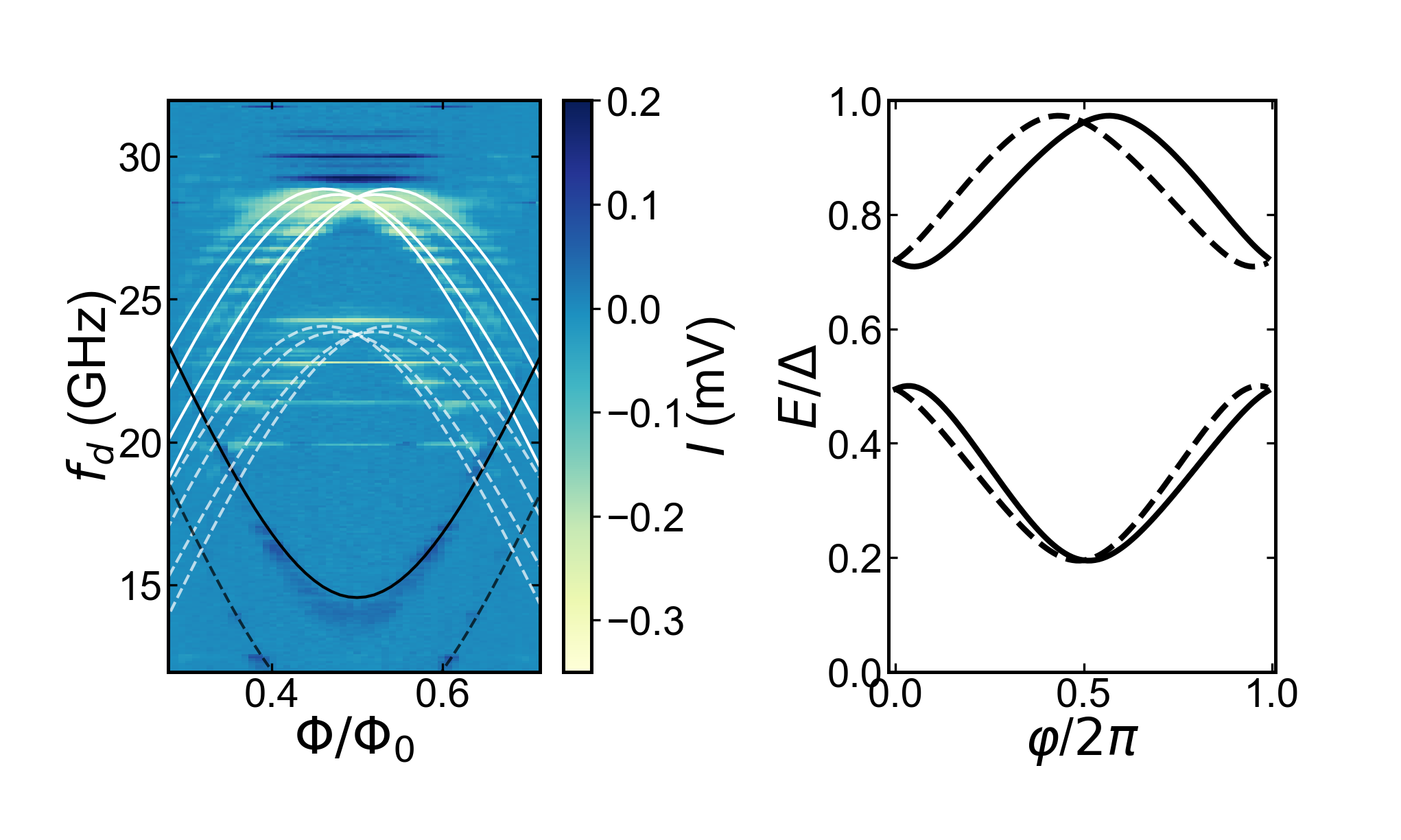}
\caption{\label{fig:sup:single-barrier-fit}  Fit of the ABS spectrum with to a single barrier model \cite{tosi_spin-orbit_2019} used to construct the energy levels of Fig.1(d). 
(a) Black and white solid lines indicate the fitted even and odd transitions and dashed lines are copies, displaced by -$f_r$ (4.82 GHz) that are visible in the data due to a finite $\langle n_\mathrm{ph}\rangle$ in the cavity during spectroscopy. The optimal single barrier model parameters are: $\Delta$=37.1 GHz, $\lambda_1$=1.37, $\lambda_2$=1.82, $\tau$=0.76, $x_r$=0.68. (b) Corresponding spin-down (solid) and spin-up (dashed) Andreev levels also shown in Fig.1 (d)}
\end{figure}
We applied the phenomenological model described in Ref.~\cite{tosi_spin-orbit_2019, park_andreev_2017} to fit a pair of even and odd transitions simultaneously. This model considers a junction with 2 sub-bands in presence of spin-orbit coupling. Only the lowest sub-band is occupied, and the lowest levels gain a spin-dependent Fermi-velocity $v_{Fj}$ due to spin-orbit interaction with the higher band. The resulting ABS energy spectrum is used in~Fig. 1(d) to illustrate the two types of transitions.

Even and odd transitions were extracted from the spectrum by thresholding \textit{I}. The fit was performed by first mapping the theoretical lines to 2D by assigning an artificial 0.2GHz wide step-function, and applying a Gaussian filter over both theory and extracted data. Finally the resulting 2D arrays are compared. The extracted model parameters are: $\Delta$=37.1 GHz, $\lambda_1$=1.37, $\lambda_2$=1.82, $\tau$=0.76, $x_r$=0.68.
Here, $\Delta$ is the superconducting gap; $\lambda_{j}$ is the ratio of the effective junction length $L$ and the ballistic coherence length, $\lambda_{j} = L/\xi = L\Delta/(\hbar v_{Fj})$, $\tau$ is the transmission probability of a single scatterer located at $x_r$ used to model a finite normal reflection probability due to elastic scattering in the junction. We refer the reader to Refs~\cite{park_andreev_2017, tosi_spin-orbit_2019} for further details about the parameters.

We are hesitant to relate these parameters to microscopic properties of the junction, because the fit was very sensitive to the initial guess and the model assumes only a single occupied sub-band, while in gate sweeps we generally see multiple ABS present (c.f~\cref{fig:sup:resonator-fit}), which can significantly distort the extracted fit parameters. However, the model shows qualitatively good agreement with the shape of the transitions shown in the data, clearly demonstrating the parity nature of the two transitions which is what is important for the conclusions drawn in this work.

\newpage
\section{Rate equations and additional fits} \label{sec:app:rate-equation}
\begin{figure*}[b]
\includegraphics{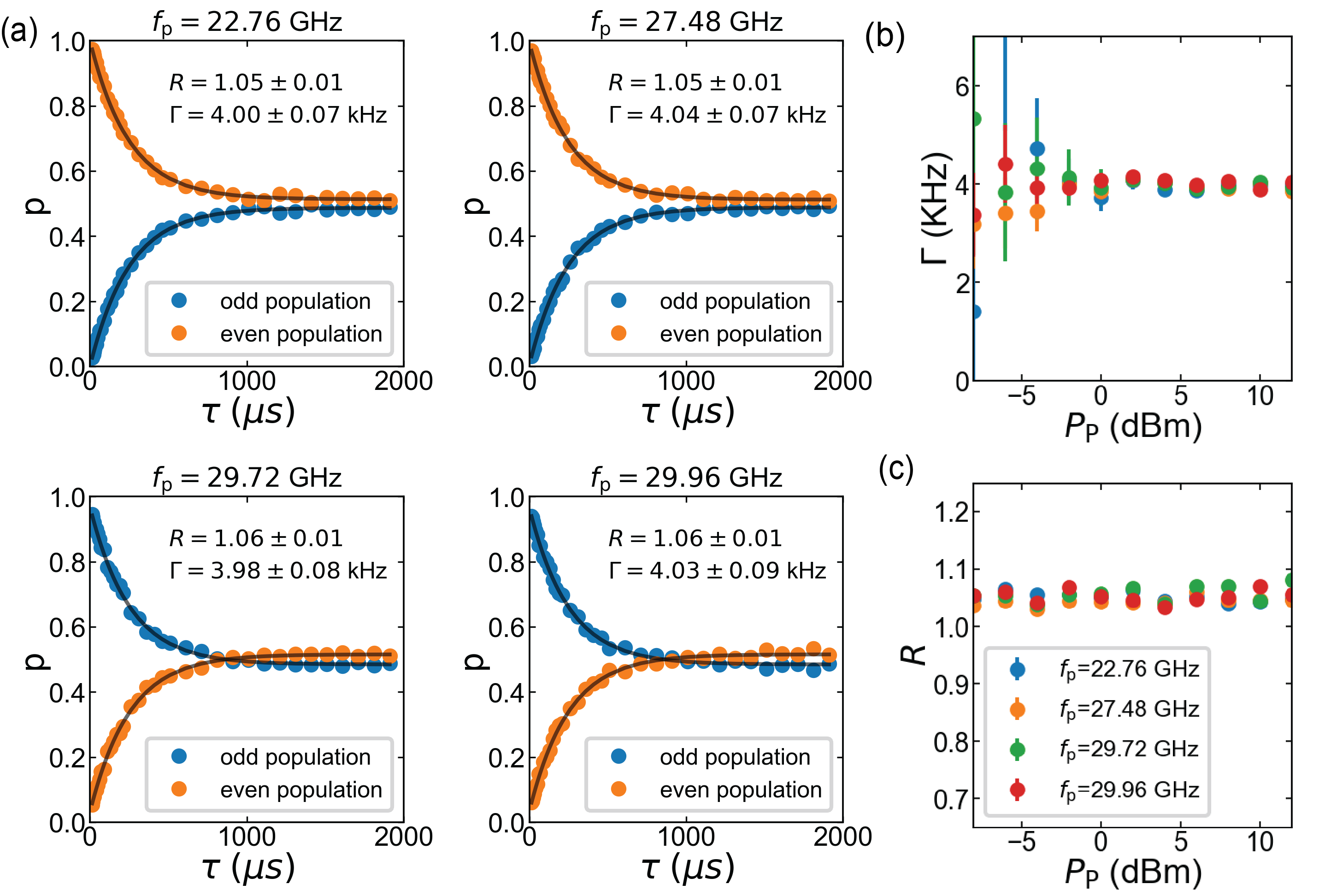}
\caption{\label{fig:sup:rate_model_undriven} (a) Fits of the population decay shown after pumping at different $f_
\mathrm{p}$ in the main text, and two additional datasets used to extract $R,\Gamma$. (b, c) extracted ratios $R$ and $\Gamma$ from the delay time fits of (a) versus $P_\mathrm{p}$, at lower pump powers the contrast goes down and the fits become more inaccurate. The fact that $R$ stays constant vs $P_\mathrm{p}$ indicates that there is no drive induced long time scale process (longer than $4 \mu s$ governing the parity imbalance (e.g a non-equilibrium QP population that remains after turning the pump off).}
\end{figure*}

The simple rate model illustrated in Fig. 3 (e) is given by
\begin{equation}\label{eq:rate_eq}
    \begin{aligned}
    \dot{p}_{\mathrm{e}} &= \Gamma_\mathrm{oe}p_{\mathrm{o}} - \Gamma_\mathrm{eo}p_{\mathrm{e}}\\
    \dot{p}_{\mathrm{o}} &= \Gamma_\mathrm{eo}p_{\mathrm{e}} - \Gamma_\mathrm{oe}p_{\mathrm{o}}\\
    \end{aligned}
\end{equation}

The general solution for a given population $p_\mathrm{e}(0)$ and $p_\mathrm{o}(0)$ at $t=0$ is given by

$$p_\mathrm{e}(t) = p_\mathrm{e}(0) + \frac{\Gamma_\mathrm{oe}p_\mathrm{o}(0)-\Gamma_\mathrm{eo}p_\mathrm{e}(0)}{\Gamma}\left(1-{\Gamma}e^{-\Gamma t}\right)$$

where $\Gamma = \Gamma_\mathrm{oe}+\Gamma_{\mathrm{eo}}$ and $p_\mathrm{o}(0) = 1-p_\mathrm{e}(0)$.  Note that we are under the (simplified) assumption that we don't have population in the excited states of each parity branch.
  We denote the populations in both spin-split levels with $p_\mathrm{o}$, since we do not resolve spin in our measurement.

\subsection{Fit of equilibrium rates}

We fit the data from Fig.3 (f) to the above model in order to extract $\Gamma$, $R$ when the drive is off. This is done by setting $t=0$ at the end of the drive pulse and then evolving the undriven rate model for a time $\tau$ in~\cref{eq:rate_eq} (adding $\SI{10}{\micro\second}$ to compensate for decay during the measurement pulse). In~\cref{fig:sup:rate_model_undriven} we show the fit results for the two frequencies used in Fig. 3 (f) of the main text, as well as for two additional pumping frequencies on which we performed the same experiment.  

Using the equilibrium values for $\Gamma$ and $R$, we can infer that the residual infidelity of the parity pumping towards even of fig. 3(d) is limited by the decay back to equilibrium during the wait time $\tau$ and the measurement pulse.
This is because evolving the equilibrium rates starting from a fully pumped $M_\mathrm{P}$ = 1 for the duration of the delay and of half the measurement pulse width (\SI{14}{\micro\second}) would give $M_\mathrm{P}$ = 0.946 (the full \SI{24}{\micro\second} would give $M_\mathrm{P}$ = 0.91).
The same explanation is not enough to explain the residual depolarization when pumping towards odd parity. This could be due to a finite power dependent pumping towards even at those frequencies, for example due to higher order odd transitions $\pm$ $f_\mathrm{r}$ occurring at high powers, since the odd spectrum is more crowded in general.

\subsection{Fit of power dependence of pumping}

In an attempt to shed light on the order of the processes involved during pumping, we now consider a modification to~\cref{eq:rate_eq} by assuming $\Gamma_\mathrm{oe},\Gamma_{eo}$ are changed during the pumping pulse.
We adopt the following phenomenological model to account for a power-dependence of the transition rates
\begin{equation}\label{eq:driven_rate}
    \left\{
    \begin{aligned}
    \mathrm{if} \,\, f_\mathrm{p}=f_\mathrm{oe}, \,\,\,\,& \Gamma_\mathrm{oe} = \Gamma_\mathrm{oe}^\mathrm{Eq}+kP_\mathrm{p}^x,\,\,\,\, &\Gamma_\mathrm{eo}=&\Gamma_{eo}^\mathrm{Eq}\\
    \mathrm{if} \,\,f_\mathrm{p}=f_\mathrm{eo}, \,\,\,\,& \Gamma_\mathrm{oe}=\Gamma_{oe}^\mathrm{Eq}\,\,\,\, &\Gamma_\mathrm{eo}=&\Gamma_\mathrm{eo}^\mathrm{Eq}+kP_\mathrm{p}^x\,\,\,\,\\
    \end{aligned}
    \right.
\end{equation}
Here, $P_p$ is the applied pumping drive power at de-trapping (trapping) frequency $f_\mathrm{oe}$~($f_\mathrm{eo}$) and $k,x$ are fitting parameters that may depend on the pump frequency. Here, $k$ is a measure of the frequency response of the circuit and tranmission lines at $f_p$ from the microwave source to the sample, which is assumed constant versus pump power. The extracted $x$ gives information on the order of the process involved during the pumping. In accordance with Fermi's golden rule, a single photon process would result in $x=1$.  $\Gamma_{oe}^\mathrm{Eq}$,$\Gamma_{eo}^\mathrm{Eq}$ represent the equilibrium rates extracted in~\cref{fig:sup:rate_model_undriven}. 
To reduce the amount of fit parameters, we assume that pumping on an even (odd) transition at $f_\mathrm{eo}$~($f_\mathrm{oe}$) only changes $\Gamma_\mathrm{eo}~(\Gamma_{oe})$.

\begin{figure*}[h]
\includegraphics{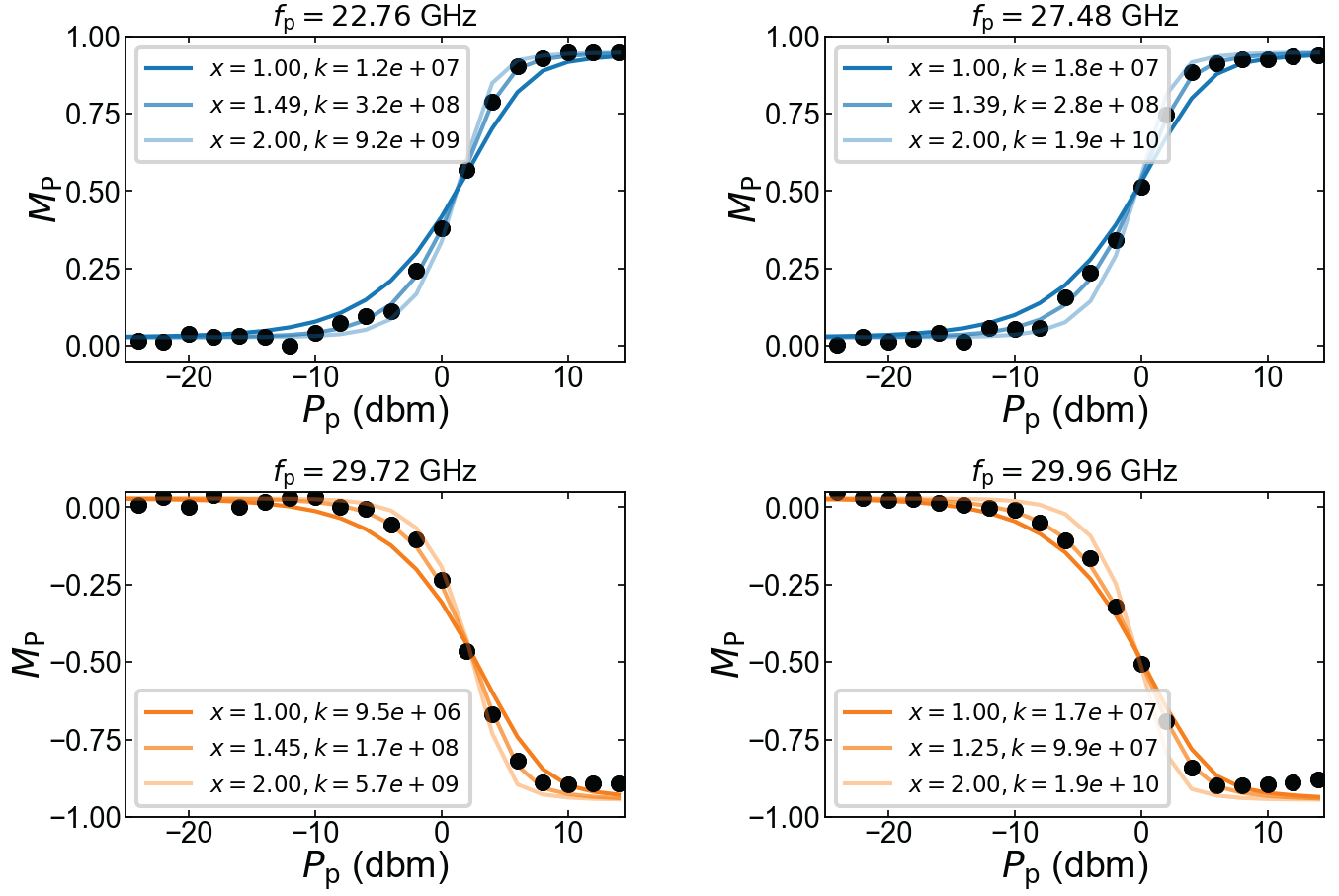}
\caption{\label{fig:sup:rate_model_driven} Power dependence at the same $f_\mathrm{p}$ as in~\cref{fig:sup:rate_model_undriven} with fits using~\cref{eq:driven_rate} with both $x, k$ as free parameters. Two additional fits are shown keeping $x$ fixed to 1,2 and fitting only $k$.}
\end{figure*}

For each power, we evolved the rate model with one of the rates made power-dependent for the duration of the pump pulse, followed by evolving the un-driven model for the wait time $\tau$ and half the measurement pulse length.
We apply this procedure to fit the power dependence at four $f_\mathrm{p}$ with $k$ and $x$ as free fit parameters~[\cref{fig:sup:rate_model_driven}]. The average of the best fit results of $x$ for the four different $f_p$ is $x=1.4\pm 0.1$.

The extracted value of $k$ varied with $f_\mathrm{p}$, because $k$ represents the absolute power as a function of frequency that arrives at the sample. Therefore, $k$ depends on the frequency response of the setup plus on-chip lines, which is not easily known from an independent measurement at frequencies outside the amplifier bandwidth.
Since $k,x$ had a large correlation coefficient in the fit, in~\cref{fig:sup:rate_model_driven} we display additional fits keeping $x$ fixed at the values indicated and fitting only $k$. 
The fact that $x=1$ doesn't fit well points towards a multiphoton nature of the polarization processes, also suggested for de-trapping in Ref.~\cite{hays_coherent_2021}.
Care has to be taken for extraction of $x$ at high powers, since from the analysis of continuously readout traces in~\cref{sec:sup:analysis_procedure_driven_traces} we found that eventually both rates start increasing, which violates one of the simplifying assumptions of the model in~\cref{eq:driven_rate}.

\subsection{Extraction of R}
By assuming the system is in equilibrium at the end of the pump pulse, we can solve~\cref{eq:rate_eq} directly: $R=1/p_\mathrm{o}-1$. This is used in Fig. 3 (g) to extract the power dependence of $R$. 
This gives a conservative estimate of $R$, because a steady state is not reached for a pump pulse length of \SI{50}{\micro\second}.  
Furthermore we neglected decay during $\tau$ and measurement time which also reduces the extracted $R$.
Note that we could have also gotten $R$ as a function of power from fitting~\cref{eq:driven_rate}, which would slightly increase the estimates shown in the main text.

\section{Effect of pump pulse length on the polarization}

\begin{figure}[ht]
\includegraphics{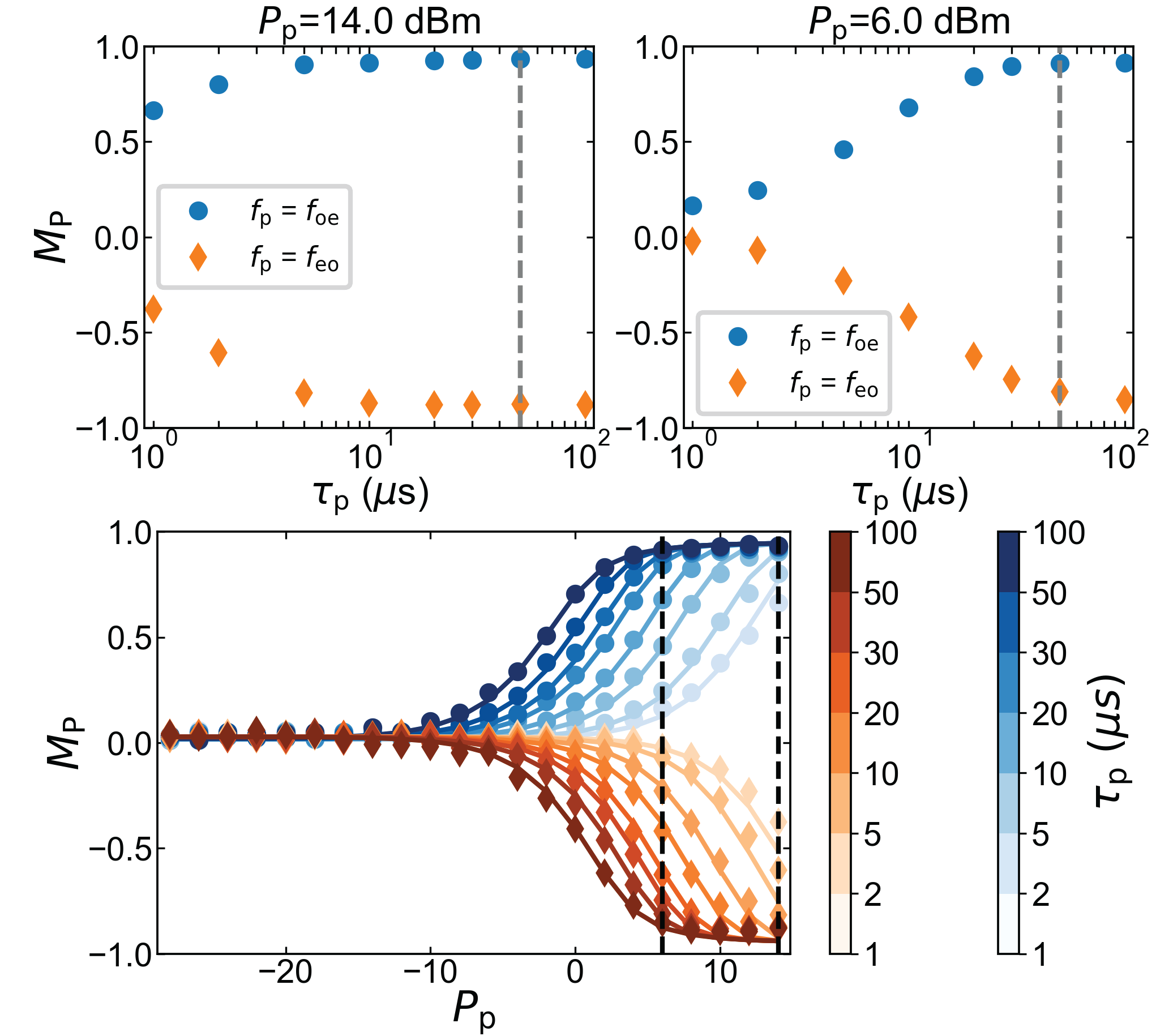}
\caption{\label{fig:sup:pump_length} Pump length dependence of the polarization. Top panels show dependence of $M_\mathrm{P}$ on the length of the pumping pulse $\tau_p$ for two pump powers $P_\mathrm{p}$ (black dashed lines in bottom panel). Grey dashed line indicates $\tau_p=\SI{50}{\micro\second}$, which is used in Fig. 3. Bottom panel indicates polarization power dependence for each $\tau_p$ (markers). Solid lines are evaluations of the driven rate model keeping all fit parameters fixed to the values obtained in~\cref{fig:sup:rate_model_driven} (for $\tau_p=\SI{50}{\micro\second}$) and only varying $\tau_\mathrm{p}$ according to the experimental setting. 
Used pump frequencies and pulse scheme were the same $f_\mathrm{p}$ as used in~Fig. 3 in the main text.}
\end{figure}

In~\cref{fig:sup:pump_length} we show how the polarization depends on the length of the pumping pulse $\tau_\mathrm{p}$ and pump power at the same pump frequencies and flux value used in the main text~Fig. 3 and~\cref{fig:sup:rate_model_driven}.
At high $P_\mathrm{p}$ we reach $M_\mathrm{P} > \pm 0.9$ already after \SI{5}{\micro\second} which could be beneficial for state-initialization protocols with high repetition rates. 

In the bottom panel~\cref{fig:sup:pump_length} we show results of the rate equation model~\cref{eq:driven_rate} for varying pump lengths $\tau_\mathrm{p}$ keeping all parameters fixed to those obtained in the fit of the \SI{50}{\micro\second} pulse in~\cref{fig:sup:rate_model_driven}. This is done both at $f_\mathrm{oe}$ and $f_\mathrm{eo}$. The agreement with the model for most $\tau_\mathrm{p}$ indicates that transient effects (a time dependent $R$ after the drive is turned on) become relevant at $\tau_\mathrm{p}<\SI{5}{\micro\second}$, where the steady state rate equation starts deviating from the data.

\newpage
\section{Parity population after readout pulse at $f_r$}

We performed a calibration experiment to make sure that the parity measurement  does not influence the populations, and therefore $M_\mathrm{p}$, at the readout amplitude used for Figs.~2 and 3. We first apply an initial $\SI{20}{\micro\second}$ readout pulse at a flux-dependent $f_r$ (using the fitting protocol described in~\cref{sec:sup:measurement_methods}) with variable amplitude $A_1$, simulating the parity readout used in the rest of this work. Then, after waiting $\SI{5}{\micro\second}$ to reset the cavity, this is followed by another $\SI{20}{\micro\second}$ pulse at $f_r$ at low amplitude $A_2=0.02$ to measure the resulting parity populations. This was repeated for multiple flux values $\Phi$. 

After rotation we fit a double Gaussian to a combined histogram of the rotated $I$ values of the 5 lowest $A_1$ (to obtain more counts and a better fit) at $\Phi=0.535\Phi_0$ where we had the largest SNR. 
Secondly, keeping $\sigma_1,\sigma_2=\sigma$ fixed to $\sigma=(\sigma_1+\sigma_2)/2 \ \forall \ \Phi$, we fitted for each $\Phi$ the means $x_1, x_2$ of again the 5 lowest $A_1$ combined. Then finally keeping all $x_1, x_2, \sigma_1, \sigma_2$ fixed to the values obtained for each $\Phi$  we fitted the amplitudes $a_1, a_2$ for each $\Phi, A_1$ value.  Note that by inspection of the fits we discarded the data at $\Phi<0.37\Phi_0$ and $\Phi>0.62\Phi_0$ since there the SNR was too low to do a proper double Gaussian fit.

\begin{figure}[ht]
\includegraphics{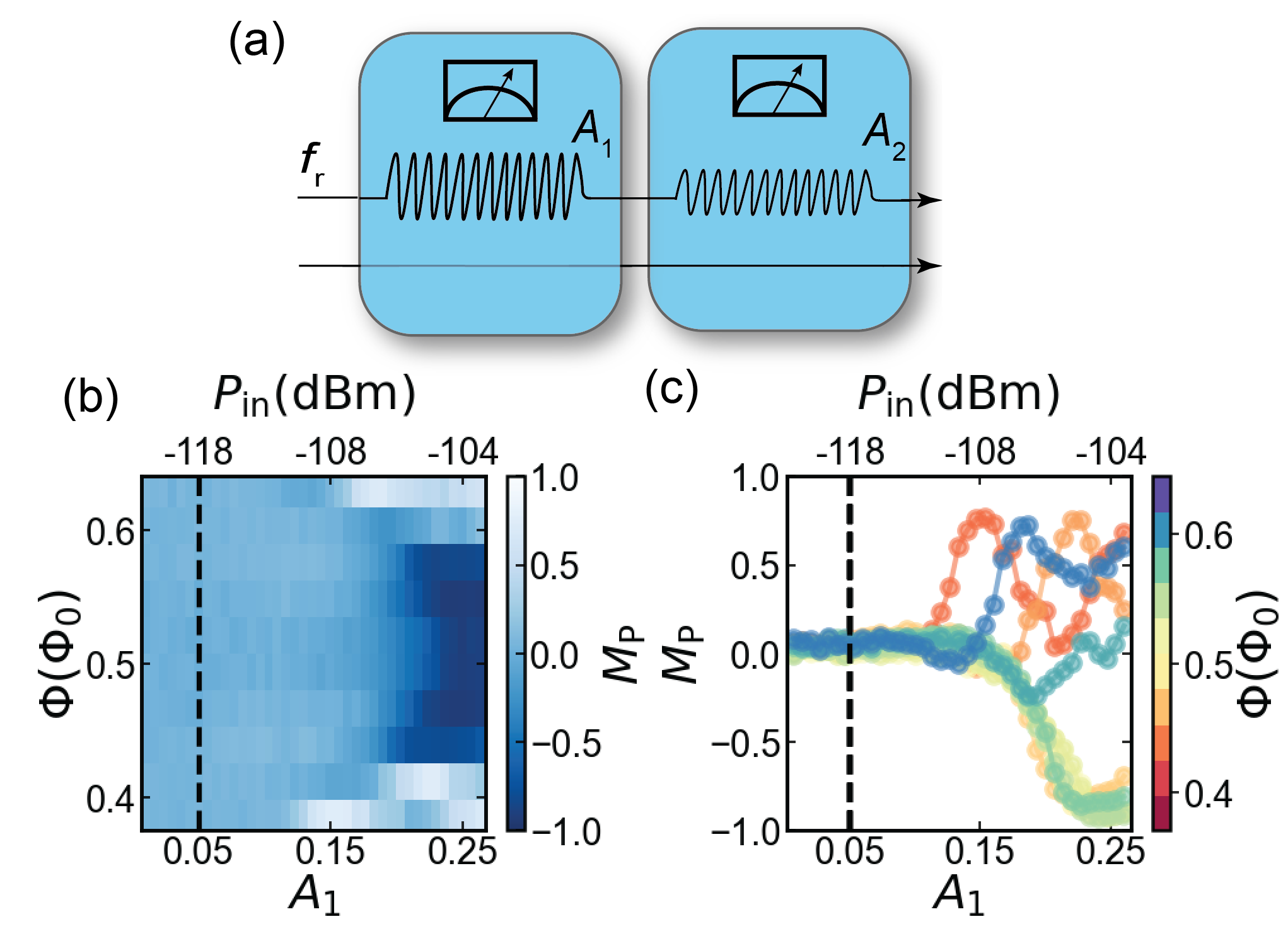}
\caption{ Effect of parity readout on the parity population. (a) Pulse scheme. A first \SI{20}{\micro\second} variable amplitude parity readout pulse is sent in, followed after waiting \SI{5}{\micro\second} by another low power $\SI{20}{\micro\second}$ parity readout pulse. (b) Population difference (odd $p_o$, minus even $p_e$) induced by the initial parity measurement resonant with the cavity frequency $f_0$ versus flux and parity pulse amplitude $A_1$ as measured by the second low power readout pulse ($A_2=0.02$). The black dashed line indicates the amplitude used for the parity readout ($A_1=0.05$) of the parity readout pulse in the rest of the paper. This is well below the values where the parity starts being pumped by a cavity tone alone.  (c) Line-cuts at different $\Phi$ (indicated in colorbar) versus $A_1$. For reference, an estimate of $L P_\mathrm{in}=V_{rms}^2/Z$, with $Z=\SI{50}{\ohm}$ and $V_\mathrm{rms}= \frac{A_1}{2\sqrt{2}}\cdot\SI {1.5}{V}$, at the input of the chip is given on the top axis. Here the attenuation $L$ includes line attenuation, known conversion losses and an additional estimated 6dB loss from the skin-effect and other sources. See~\cref{fig:sup:wiring-diagram}. In the region of the oscillations at high $A_1$ the response of the cavity (when inspecting the first pulse I-Q outcomes) becomes highly non-linear which makes a clear interpretation challenging.}\label{fig:parity_proof}
\end{figure}

The result of the second pulse parity measurement is shown in~\cref{fig:parity_proof}. At the amplitude $A_1=0.05$ used for the parity measurements of the main text, the populations are not affected by the parity measurement itself.
However, at higher readout power, the parity can be pumped by the readout tone alone, as also found in previous works~\cite{janvier_coherent_2016}. 
Note that the pump direction due to photons in the cavity switches sign around 0.43$
\Phi_0$ and 0.57$\Phi_0$, pumping towards even instead of odd parity.
This feature is not fully understood: it could be related to the change of ABS transition frequencies with flux relative to other resonances coupled to the cavity, or to multi-photon transitions involving the cavity~\cite{olivares_dynamics_2014}.

\newpage
\section{Readout power dependence of pulsed pumping process}\label{sec:sup:discussion-power-dependence}
\begin{figure}[ht]
\includegraphics{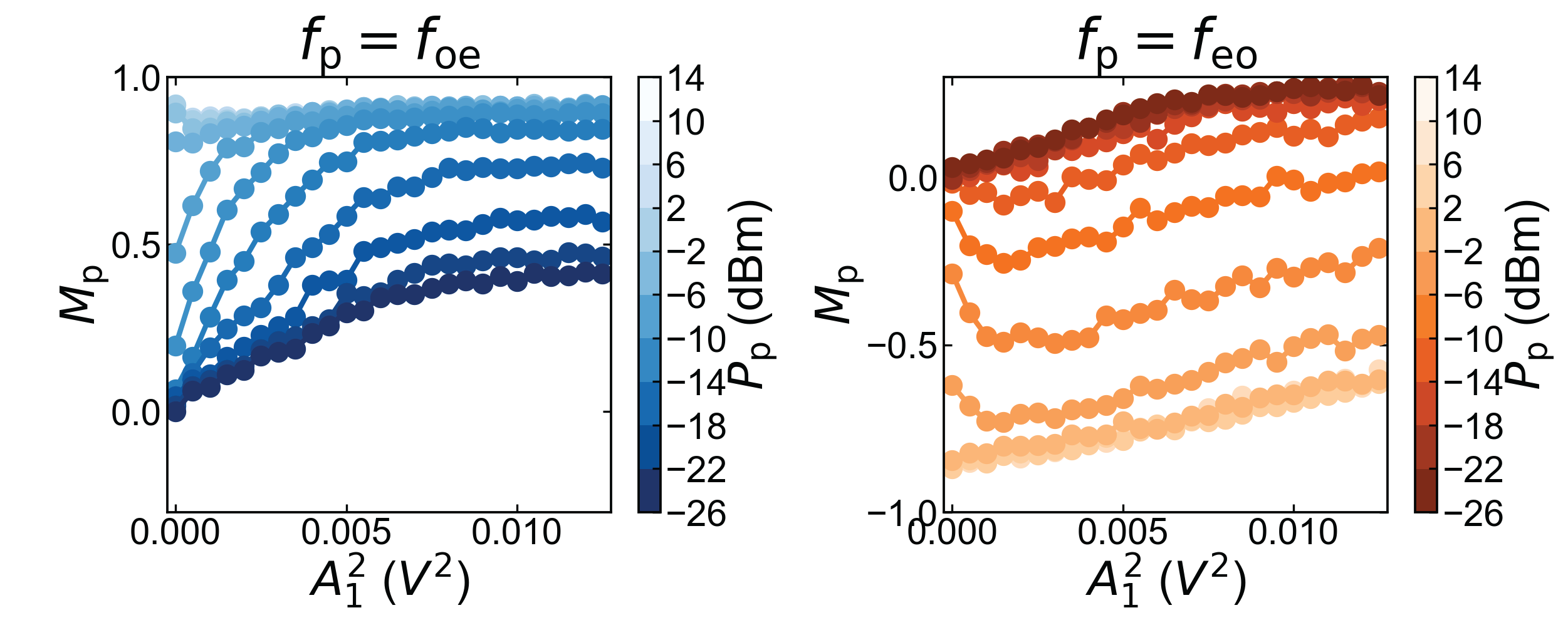}
\caption{ Pump versus readout power dependence of the pumping pulse at $\Phi=0.44\Phi_0$. The pulse sequence of~\cref{fig:sup:pulse_sequence_3c} was used. Depicted is $M_\mathrm{P}$ versus squared readout amplitude $A_1$ of the tone at $f_r$ and $P_\mathrm{p}$ of the tone at $f_p$ for $f_p=f_\mathrm{oe}=27.48$ GHz (left graph) and $f_p=f_\mathrm{eo}=29.72$ GHz (right graph) for different $P_\mathrm{p}$ as indicated on the colorbar.}\label{fig:pump_vs_readout_power}
\end{figure}

The pumping sequence of Fig. 4 and Fig. 3(c) had a weak ($A_1=0.02$) cavity tone on during the pumping, since we assumed that would facilitate multi-photon transitions towards the continuum.
We investigate the effect of pumping parity when a second tone at $f_r$ is present in~\cref{fig:pump_vs_readout_power}. The pulse sequence of~\cref{fig:sup:pulse_sequence_3c} was used. We then varied the amplitude of the first pulse tone at $f_r$ as well as $P_\mathrm{p}$. 
In~\cref{fig:parity_proof} we already found that a readout tone can polarize the parity by itself, where the polarization direction depends on the applied phase bias.  
The results of ~\cref{fig:pump_vs_readout_power} show that indeed for lower pump powers a weak cavity tone helps the pumping process (in both directions). However, at strong pump power the highest polarization is actually achieved for $A_1=0$ in both pump directions. A possible explanation could be that with increasing $\langle n_\mathrm{ph} \rangle$ in the cavity the total $\Gamma$ increases (as seen in~\cref{fig:rates_even}), reducing $R$ effectively. A general trend of pumping towards even parity with $A_1$ is also visible.

\newpage
\section{Continuous readout during pumping}
\label{sec:sup:analysis_procedure_driven_traces}

\begin{figure*}[b]
    \centering
    \includegraphics{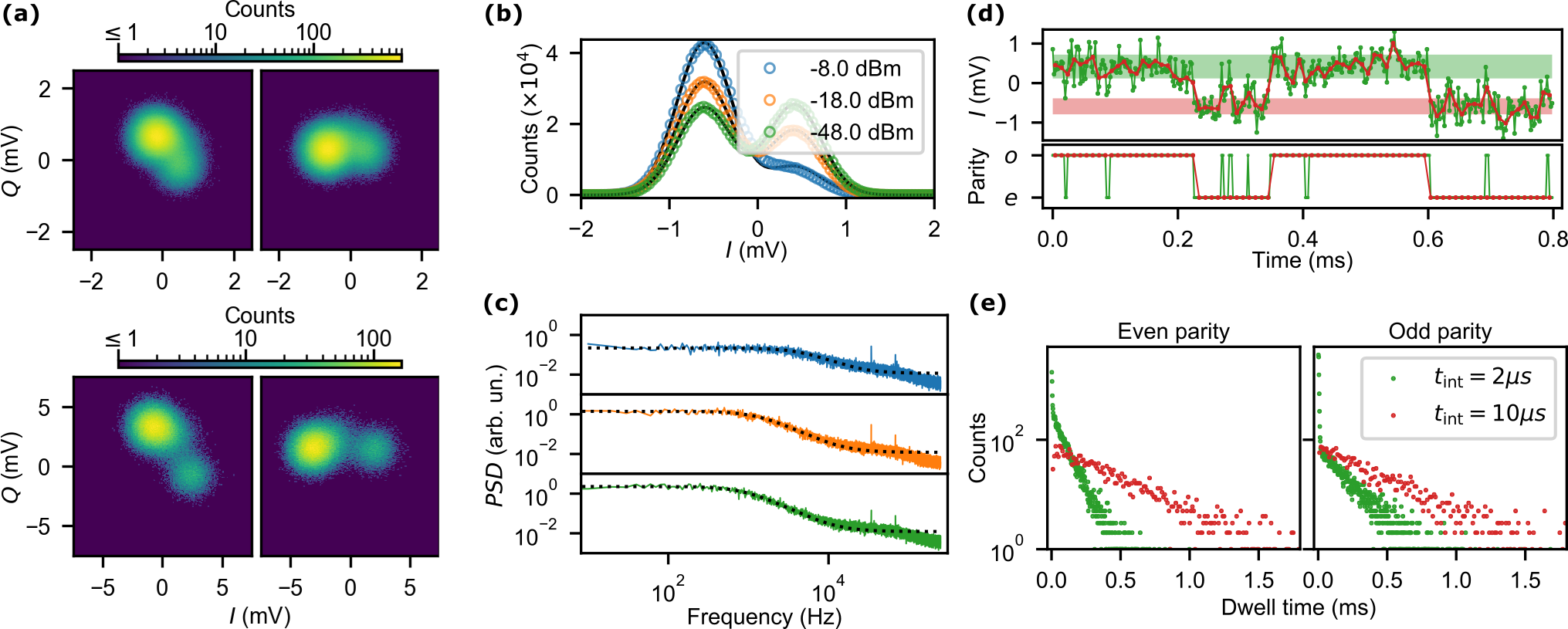}
    \caption{Analysis procedure for driven parity switching traces. \textbf{(a)} 2D histogram of the raw (left) and rotated (right) measured data in the $IQ$ plane to project parity information solely into the $I$ quadrature. The top panel shows the original data with $t_\mathrm{int} = \SI{2}{\micro\second}$, while the bottom panel shows the data obtained by summing five consecutive points. \textbf{(b)} Histogram of the $I$ quadrature of rotated raw measurement data (points) and fit of a double Gaussian distribution (solid black line) for three different drive tone powers. At high drive power the amplitude of one Gaussian decreases, indicating the decreasing presence of the associated junction parity. \textbf{(c)} Power spectral density of the projection to the $I$ quadrature of the rotated time resolved data (colored lines, legend of panel (c) applies) and fit of a Lorentzian (dashed line) yielding a characteristic transition rate~\cite{Machlup1954}. \textbf{(d)} Time resolved $I$ quadrature projection of the rotated data (top) and corresponding state assigned using a two-point filter \cite{Vool2014}. Green (red) points indicate the data for \SI{2}{\micro\second} (\SI{10}{\micro\second}) integration time. Shaded areas indicate $\pm 1\, \sigma$. \textbf{(f)} Histogram of dwell times in the even and odd parity for \SI{2}{\micro\second} (green) and \SI{10}{\micro\second} (red) integration time. In the distributions should be identical and single exponential for both integration times assuming sufficient SNR, and a purely Poissonian switching process. We attribute the double exponential distribution for short integration time to a finite overlap between the two Gaussians, i.e. too low SNR.\label{fig:td_traces}}
\end{figure*}

As an alternative verification of the parity pumping process, we perform experiments with continuous driving and readout of the ABS (two traces shown in Fig.3 (a)), similar to e.g. Ref.~\cite{janvier_coherent_2015, hays_direct_2018}. Opposed to the pulsed experiments described before, we now send a continuous microwave tone at fixed frequency and power to readout and drive line, respectively, and record traces of \SI{2}{\second} for various combinations of drive frequency $f_\mathrm{d}$, drive power $P_\mathrm{d}$ and readout amplitude $A_{\mathrm{ro}}$. The experiments were performed at the same $f_\mathrm{p}$ and a flux $\Phi=0.46$ close to the pulsed experiments in order to pump on the same transitions. After down-conversion and demodulation we integrate the  signal for \SI{2}{\micro\second} per point and store $10^6$ points per trace. Given the relatively slow equilibrium parity switching rates, we sum five consecutive points from the original raw data to increase separation between the two clusters of points, i.e. increase SNR, while sacrificing time resolution with an effective integration time of $t_\mathrm{int} = \SI{10}{\micro\second}$ [see top vs. bottom panel of~\cref{fig:td_traces}(a)].
We rotate the time series of points in the $IQ$ plane such that we achieve maximum contrast in the $I$ quadrature[cf. right panels of~\cref{fig:td_traces}(a)]. Following this we fit again a double Gaussian distribution to a histogram of the $I$ values of the rotated data. 

To extract the characteristic transition rates between even and odd parity we obtain the power spectral density (PSD) of the time series $I(t)$ extracted from the rotated complex data by fast-Fourier transformation [cf. Fig.\ref{fig:td_traces}(c)]. To reduce the noise in the PSD, we take the original $2\times 10^5$ samples long time trace and reshape it into 20 non-overlapping segments of equal length, finally we average the 20 PSD obtained from the individual segments \cite{bartlett_smoothing_1948}. We fit the averaged PSD of a random telegraph switching process with two characteristic rates $\Gamma = \Gamma_\mathrm{oe} + \Gamma_\mathrm{eo}$ \cite{Machlup1954} $PSD(\omega) = a \, 4 \Gamma / \left(\Gamma^2 + \omega^2\right) + c$, where $c$ accounts for constant background noise. By assuming a two state rate equation model in steady state, we are finally able to extract the individual parity switching rates from the fitted values of $\Gamma = \Gamma_\mathrm{oe} + \Gamma_\mathrm{eo}$ and the fitted Gaussian amplitudes $a_1 / a_2 = \Gamma_\mathrm{oe} / \Gamma_{eo}$. 

To check the underlying assumption of the analysis outlined above, namely uncorrelated parity switching events, we also analyze the recorded $I(t)$ directly in the time domain by applying a two-point filter \cite{Vool2014}. \cref{fig:td_traces} (d) illustrates the raw recorded time traces for \SI{2}{\micro\second} integration time for a drive power of \SI{-48}{\dBm} in green, with the green shaded area indicating $\pm 1 \sigma$ of the Gaussian histogram of all data points [cf. \cref{fig:td_traces} (b)]. Red data points indicate the average of 5 consecutive raw points similar to~\cref{fig:td_traces}(a).
The lower panel of~\cref{fig:td_traces}(d) shows the parity assigned by the two point filter in the color corresponding to the data on which it is based. If a Poisson process governs the parity switches the histogram of the dwell times in even and odd parity should show an exponential distribution. Note, however, that non-Poissonian quasiparticle processes have been observed \cite{Vool2014} and could in principle also be present in the device investigated in this paper.
\cref{fig:td_traces}(e) shows typical histograms of the dwell times in even and odd parity extracted from the state assignment by the two-point filter for 2 (green) and \SI{10}{\micro\second} (red) integration time. For short integration time, we observe a large excess count of short dwell times. We attribute this to be an artifact of the limited SNR. By increasing the integration time, and consequently also SNR, the excess counts of short dwell times vanish and we recover exponential distributions of the dwell times in even and odd parity as expected for Poissonian processes.

By fitting an exponential distribution to the dwell time histograms we extract the characteristic transition rates for even and odd parity directly from the time series.
We observe good agreement between PSD and two-point filter method for low drive powers and rates that are much slower than $1/t_\mathrm{int}$. However, for increasingly fast transition rates the corresponding histogram of dwell times has a rapidly decreasing number of points making the fit of the exponential distribution unreliable. For consistency, we therefore use the PSD method for all analysis presented in the following sections. 

\subsection{Power dependence of transition rates}
\label{sec:sup:pwr_dep_driven_traces}

\begin{figure*}[b]
    \centering
    \includegraphics{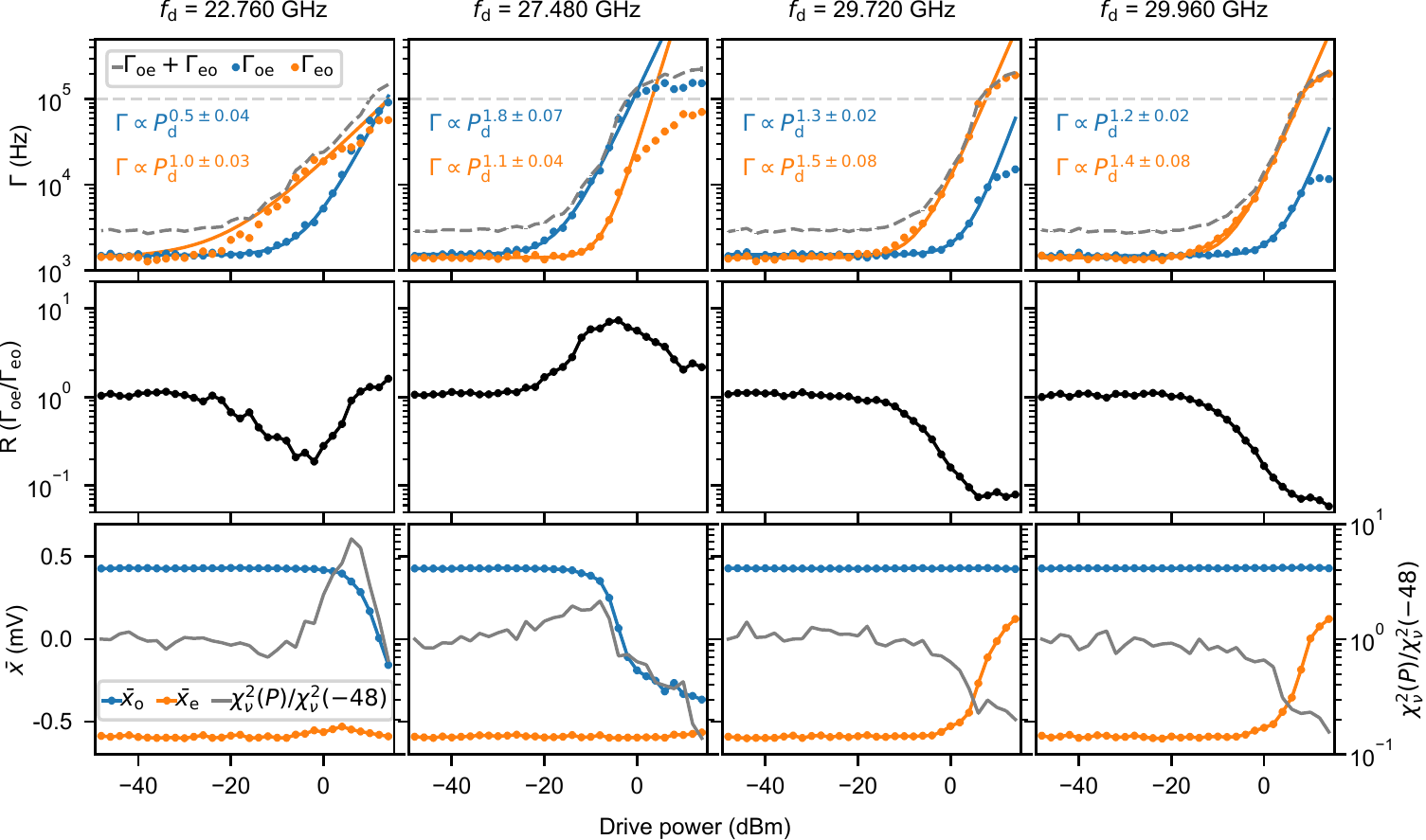}
    \caption{  Parity switching rates as a function of drive power at frequencies indicated at $\Phi=0.46\Phi_0$ . \textbf{Top} Parity transition rates as a function of drive power (blue, orange marker), and sum of both rates (grey dashed line). Colored lines are fits of the corresponding data to Eqn.~\ref{gen_ratemodel} with fitted exponents given in the respective panels. The horizontal grey dashed line indicates $1/t_\mathrm{int}$, roughly the maximum resolvable transition rate. \textbf{Middle} Ratio of $\Gamma_\mathrm{eo}/\Gamma_\mathrm{oe}$ as a function of drive power. \textbf{Bottom} Means of the two Gaussian distributions indicating even (orange) and odd (blue) parity. As the transition rate approaches $1/t_\mathrm{int}$ the mean of the Gaussian associated with the pumped parity moves towards the other one, and the normalized reduced $\chi^2$ (grey line, right y-axis) deviates strongly, indicating that the goodness of fit decreases due to approaching the limit of the experimental time resolution.}\label{fig:rates_odd}
  
\end{figure*}
Similarly to the analysis of the pulsed measurements in Fig.3 of the main text, we extract parity switching rates as a function of drive power. The top row of~\cref{fig:rates_odd} shows the transition rates between even (orange) and odd (blue) parity for the four different driving frequencies as a function of drive tone power.
The markers indicate the rates obtained following the PSD approach (cf.~\cref{sec:sup:analysis_procedure_driven_traces})  using $t_\mathrm{int} = \SI{10}{\micro\second}$. A light-gray dashed line indicates $1/t_\mathrm{int}$ to show where the extracted rate becomes comparable to the time resolution of the measurement, and the sum of both rates is indicated by a dark grey dashed line. We fit the obtained rates ($\Gamma_\mathrm{oe},\Gamma_\mathrm{eo}$) using a generic model
\begin{equation}
\label{gen_ratemodel}
\Gamma(P) = \Gamma_0 + k\, P^x,    
\end{equation}
where the power $P$ is given in Watt. The top row of~\cref{fig:rates_odd} shows the rates together with the best fit curves. 

Different exponents for the different driving frequencies, and onsets of the rate change could be either due to the underlying physical process, or a due to the frequency dependent transmission of the drive line. For high drive powers, the transition rates surpass the time resolution $\approx 1/t_\mathrm{int}$, and the observed flattening is likely an artifact of this fact. We show the ratio $R = \Gamma_\mathrm{oe}/\Gamma_\mathrm{eo}$ in the middle row of~\cref{fig:rates_odd}, and observe a power dependent change in the ratio up to a factor $\approx 10$. Finally, the bottom row indicates the mean value of the two Gaussians forming the double Gaussian distribution of the measurement results in the $IQ$-plane. As can be seen, the Gaussian indicating the parity we are dynamically polarizing to stays constant, while the mean position of the parity polarized away from moves towards the former. Additionally, we observe a decrease proportional to $R$ in the pumped parity Gaussian's amplitude. Finally, due to the increased transition rates between the two parities, the Gaussian we are polarizing away smears out and gradually merges into the Gaussian indicating the dynamically polarized parity.
For $f_p=\SI{22.76}{\giga\hertz}$ we see an opposite trend compared to~\cref{fig:sup:rate_model_driven} ( $\Gamma_\mathrm{eo}$ increases first while~\cref{fig:sup:rate_model_driven} shows an increase of $\Gamma_\mathrm{oe}$). We attribute this to the $0.02\Phi_0$ difference in flux setting causing a move off resonance with the odd transition. This is not the case for the other $f_\mathrm{p}$ (see Fig.3 (c) for the mapping).

\begin{figure*}
    \centering
    \includegraphics{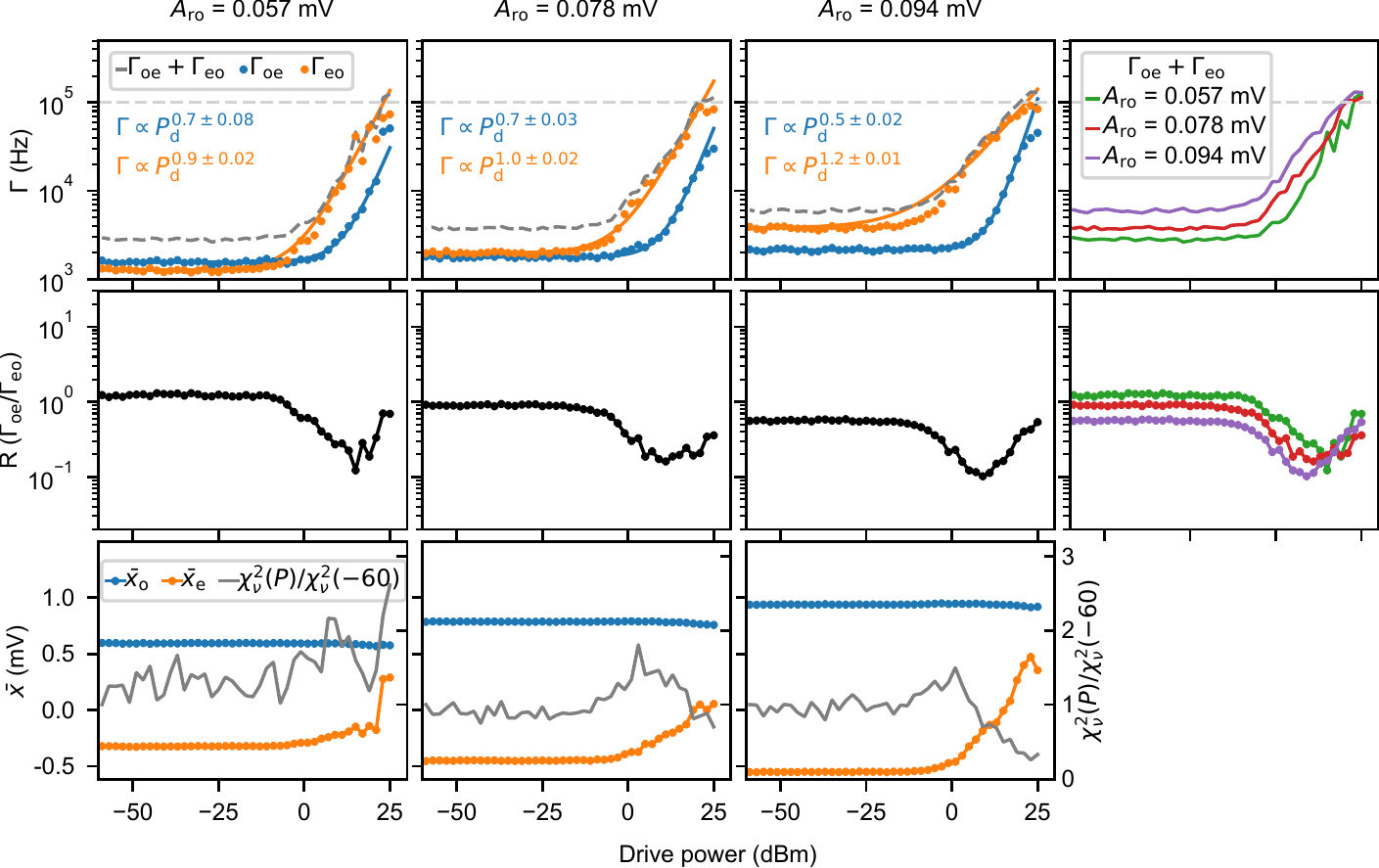}
    \caption{Parity switching rate as a function of drive power at $f_d = \SI{17.5}{\giga\hertz}$ resonant with the lowest available even pair transition for three different readout powers. Note that there was \SI{6}{\deci\bel} less attenuation on the drive line compared to the drive-power axes of all previously presented data. \textbf{Top} Parity transition rates as a function of drive power (blue, orange markers), and sum of both rates (grey dashed line). Colored lines are fits of the corresponding data to~\cref{gen_ratemodel} with fitted exponents given in the respective panels. The horizontal grey dashed line indicates $1/t_\mathrm{int}$, the maximum resolvable transition rate. The right most panel compares the total rates as a function of drive power for the different readout amplitudes and indicates an increase of the rates with increasing readout power. \textbf{Middle} $R$ as a function of drive power. The right most panel compares the drive power dependend ratios for the three different readout amplitudes (same legend as in the top row applies). \textbf{Bottom} Means of the two Gaussian distributions indicating even (orange) and odd (blue) parity. As the transition rate approaches $1/t_\mathrm{int}$ the mean of the Gaussian associated with the pumped parity moves towards the other one, and the normalized reduced $\chi^2$ (grey line, right y-axis) deviates, indicating that the goodness of fit decreases due to approaching the limit of the experimental time resolution.}\label{fig:rates_even} 
\end{figure*}

\cref{fig:rates_even} shows the power dependence of transition rates between even and odd parity driving on resonance with the lowest available even transition $f_\mathrm{d} = \SI{17.5}{\giga\hertz} ,\Phi=0.60$, for three different readout amplitudes $A_\mathrm{ro}$ applied at $f_\mathrm{r}$.
Note that, compared to driving a higher frequency even transition [cf.~\cref{fig:rates_odd}], the fitted exponent is lower here, while the onset of pumping starts $\sim$\SI{20}{\deci\bel} higher. Since the drive frequency we are using here is lower, we would expect a higher order process, which is consistent with the larger power needed for the onset of pumping, but inconsistent with the smaller fitted exponent.
Increasing the readout amplitude by about a factor of two results in $\sim 3$ times larger switching rate from even to odd parity (orange dots). We hypothesize this is due to effective parity pumping by the readout tone [cf. \cref{fig:parity_proof}].

For all three readout amplitudes, the ratio between the parity transition rates follows a similar trend (see middle plot in last column of~\cref{fig:rates_even}), and decreases by about an order of magnitude. For even higher powers $R$ increases again until the rate extraction becomes uncertain due to $\Gamma \sim 1/t_\mathrm{int}$.  Similar to the bottom row of~\cref{fig:rates_odd}, the bottom row of~\cref{fig:rates_even} shows the means of both Gaussians, which constitute the double Gaussian distribution indicating the two parities. For drive powers $> \SI{0}{\deci\bel}$ the Gaussian associated with even parity moves towards the constant mean of the odd parity Gaussian. 

In summary, the continuously measured traces support the conclusions as presented with the pulsed experiments. Here we can obtain both rates separately when the drive is on. This shows that with stronger readout amplitude as well as with strong drive power, both rates increase.  However, the analysis does not capture excited ABS populations which are present (the driven blob starts spreading outward in ~\cref{fig:td_traces}). At high drive powers possible distortions of the readout signal due to the strong drive tone come into play as well.  This is why we applied a pulsed scheme that avoids these caveats to support the main conclusions of this work. Future work could include excited populations in the model for the jump traces, which we did not attempt here because the short coherence times relative to our SNR did not allow for a clear separation of the excited populations from their parity ground state. 
\bibliography{bibliography2}